\documentclass[reprint,pre,aps,  nolongbibliography
	footinbib
	]{revtex4-2}
 
\usepackage{geometry}
\usepackage{todonotes}
\usepackage{amsmath,amsthm,amssymb, mathtools,mathrsfs}
\usepackage{physics}
\usepackage{amsfonts, latexsym, amssymb}
\usepackage{bbm, dsfont}
\usepackage{caption}
\usepackage{graphicx}
\usepackage{float}
\usepackage{url}
\usepackage{xcolor}
\usepackage{hyperref}
\usepackage{tikz}
\usetikzlibrary{decorations.pathmorphing,cd,decorations.markings,calc,fadings}
\hypersetup{
    colorlinks=true,
    linkcolor=blue,
    filecolor=magenta,      
    urlcolor=cyan,
    citecolor=magenta,
}
\usepackage{enumerate}
\usepackage{bm}

\definecolor{darkgreen}{rgb}{0.1,0.6,0.1}

\newcommand{\nullify}[1]{}

\newcommand{\be}{\begin{equation}} \newcommand{\ee}{\end{equation}}
\newcommand{\ben}{\begin{equation*}} \newcommand{\een}{\end{equation*}}

\newcommand{\bea}{\begin{equation} \begin{aligned}} \newcommand{\eea}{\end{aligned} \end{equation}}

\def\repa{\raise4pt\hbox{$\square$}\mkern-14mu\raise-4pt\hbox{$\square$}}
\def\repab{\overline{\raise4pt\hbox{$\square$}\mkern-14mu\raise-4pt\hbox{$\square$}\mkern-1mu}}

\usepackage{tikz}
\usetikzlibrary{shapes.misc}
\tikzset{cross/.style={cross out, draw=black, ultra thick, minimum size=2.3*(#1-\pgflinewidth), inner sep=0pt, outer sep=0pt},
cross/.default={5pt}}

\usetikzlibrary{decorations.pathmorphing}

\tikzset{snake it/.style={decorate, decoration=snake}}

\begin{document}

\title{Worldsheet for Generalized Veneziano Amplitudes
}
\author{Shota Komatsu$^{a}$ and Pronobesh Maity$^{b}$}
\affiliation{${}^{a}$Department of Theoretical Physics, CERN, 1211 Meyrin, Switzerland}
\affiliation{${}^{b}$Laboratory for Theoretical Fundamental Physics, EPFL, Rte de la Sorge, Lausanne, Switzerland}
\email{shota.komatsu@cern.ch}
\email{pronobesh.maity@epfl.ch}

\date{\today}

\begin{abstract}
    We present a worldsheet action that reproduces a class of dual resonance amplitudes discussed in the literature, which generalize the Veneziano amplitude for open strings. Our proposal builds on the chiral composite linear dilaton introduced recently. We further compute higher-point extensions and closed-string analogs, which exhibit partial crossing symmetry.
\end{abstract}

\maketitle
\section{Introduction}
In 1968, Veneziano proposed \cite{Veneziano:1968yb} a mathematical expression for a $2\to 2$ scattering amplitude of mesons that incorporates both the $s$-channel and $t$-channel resonances into a single analytic function, exhibiting the {\it dual resonance property}. This inspired numerous generalizations and variants \cite{Mandelstam:1968czc, Bardakci:1968rse, Coon:1969yw, Matsuda:1969zza, Chan:1969ex, Virasoro:1969me,  Gross:1969db, Shapiro:1970gy, Romans:1989di, Fairlie:1994ad}, all sharing the same feature: simultaneous encoding of different scattering channels in one expression.

With the advent of quantum chromodynamics (QCD), these ideas lost prominence and came to be seen merely as historical precursors to string theory. However interest in these amplitudes has recently revived in the context of the S-matrix bootstrap \cite{Huang:2020nqy, Ridkokasha:2020epy,  Maity:2021obe, Arkani-Hamed:2022gsa,  Figueroa:2022onw,  Huang:2022mdb,  Geiser:2022icl, Chakravarty:2022vrp,   Geiser:2022exp, Bhardwaj:2022lbz, Chen:2022shl,   Cheung:2023adk, Chen:2023dcx, Jepsen:2023sia, Cheung:2023uwn, Chiang:2023quf, Haring:2023zwu, Bhardwaj:2023eus, Rigatos:2023asb, Berman:2023jys, Geiser:2023qqq,  Jepsen:2023sia, Eckner:2024ggx, Rigatos:2024beq, Eckner:2024pqt, Wang:2024wcc, Berman:2024wyt,      Arkani-Hamed:2024nzc, Berman:2024wyt, Bhardwaj:2024klc, Mansfield:2024wjc, Bhat:2024agd, Gadde:2025fil} , motivated by two main questions. First is the {\it uniqueness problem} \cite{Caron-Huot:2016icg,Cheung:2022mkw,Arkani-Hamed:2023jwn,Albert:2024yap,Cheung:2024uhn,Cheung:2024obl,Cheung:2025tbr}: do basic principles such as unitarity, crossing symmetry, and Regge behavior uniquely determine string theory amplitudes, or are there other consistent solutions? Second is {\it the search for amplitudes} that could describe scattering in large $N$ QCD, capturing features such as confinement and Regge trajectories \cite{ Albert:2022oes, Albert:2023jtd, Fernandez:2022kzi,Albert:2023seb}.

Despite extensive early works, a worldsheet description exists only for the original Veneziano amplitude. The absence of underlying worldsheet formulations has hindered progress in systematically studying higher-point generalizations and in understanding the origins of their physical and analytic properties including its factorizability. 

In this paper, we propose a worldsheet action for a class of generalized Veneziano amplitudes introduced by Mandelstam \cite{Mandelstam:1968czc}, which depends on three parameters \footnote{We set the parameter $a$ in \cite{Mandelstam:1968czc} to $1$ by rescaling $s$ and $t$. We also introduced a new parameter $\lambda$ following \cite{Haring:2023zwu}.}, $b$, $\lambda$ and $\delta$:
\begin{align}\label{Mandelstam}
\begin{aligned}
&T(s,t)=\\
&\int_{0}^{1}dz\, z^{-s-b-1}(1-z)^{-t-b-1} \, (1-4\lambda z(1-z))^{\delta}\\
&= \frac{\Gamma(-s-b)\Gamma(-t-b)}{\Gamma(-s-t-2b)} {}_3F_2\left[\substack{-s-b,-t-b,-\delta\\ -\frac{s+t+2b}{2}, \frac{1-(s+t+2b)}{2}};\lambda\right] \, .
\end{aligned}
\end{align}
For $\lambda=b=0$ or $\delta=b=0$, this reduces to (the polarization-independent part of) the Veneziano amplitude while this agrees with the  open bosonic string amplitude upon setting $b=1$ and $\delta=0$. Our construction uses the {\it chiral composite linear dilaton theory} (CLD) introduced in our earlier work \cite{Komatsu:2025sqo}. Using this framework, we also compute higher-point amplitudes and closed-string analogs. Unlike the open-string four-point case, these extensions are not fully crossing symmetric due to hidden quantum numbers carried by the strings. Nonetheless they can be made symmetric in several specific channels, retaining part of the dual resonance structure.

\section{Generalized Veneziano amplitudes from worldsheet}
Here we present a worldsheet theory reproducing the generalized Veneziano amplitudes \eqref{Mandelstam}. We first provide a minimal setup that corresponds to a specific choice of the parameters $b$, $\lambda$ and $\delta$ and then discuss generalizations.
\vspace{10pt}

\noindent\textbf{Action and stress tensor.} 
Consider the following action for bosonic open string,
\begin{align}\label{action_1}
S=S_{\mathbb{R}^{d-1,1}}+S_{\beta\gamma}+S_{\chi} +S_{bc}\,,
\end{align}
where $S_{\mathbb{R}^{d-1,1}}$ is an action of $\mathbb{R}^{d-1,1}$ free bosons
\begin{align}
S_{\mathbb{R}^{d-1,1}}=\int_{\Sigma}\frac{d^2z}{2\pi 
}\partial X^{\mu}\bar{\partial}X_{\mu}\,,
\end{align}
and $S_{bc}$ is the action for $bc$ ghosts. Throughout this letter, we set $\alpha'=1$ and $\partial:=\partial_z$ and $\bar{\partial}:=\partial_{\bar{z}}$. $S_{\chi}$ is an arbitrary CFT with the central charge $c_{\chi}$, which describes the dynamics on an internal manifold. The most important part of the action is
$S_{\beta\gamma}$, the action for CLD \cite{Komatsu:2025sqo, Komatsu:2025dqv} given by
\begin{align}\label{CLD}
\begin{aligned}
S_{\beta\gamma}=&\int_{\Sigma}\frac{d^2z}{2\pi} \left[\beta\bar{\partial}\gamma+\bar{\beta}\partial\bar{\gamma}+q\left(\partial\varphi \bar{\partial}\varphi+\frac{\hat{R}}{2}\varphi\right)\right]\\
&+\frac{q}{\pi}\int_{\partial \Sigma} ds \, \hat{k} \varphi\,,
\end{aligned}
\end{align}
where $\varphi \equiv \log (\partial \gamma \bar{\partial} \bar{\gamma}/R^2)$, and $\hat{R}$ and $\hat{k}$ are the Ricci scalar and the geodesic curvature on the worldsheet respectively. $q$ is a $c$-number parameter, which is fixed uniquely by requiring the conformal anomaly cancellation as shown below in \eqref{eq:anomaly}. Unlike the $\beta\gamma$-system that appears in superstring theory, $\beta$ here is a $(1,0)$-form while $\gamma$ is a $(0,0)$-form and should be regarded as a target space coordinate of an extra ``internal" manifold. In what follows, we compactify $\gamma$ on a torus, by identifying the real part  $\gamma^{1}\equiv (\gamma+\bar{\gamma})/2$ as $\gamma^{1}\sim \gamma^{1} +2\pi R$ and the imaginary part $\gamma^{2}\equiv (\gamma-\bar{\gamma})/2i$ as $\gamma^2 \sim \gamma^2+2\pi \tilde{R}$. 


A novelty as compared to our previous work \cite{Komatsu:2025sqo,Komatsu:2025dqv} is 
the second line of \eqref{CLD}, which is required for the Weyl invariance in the presence of worldsheet boundaries $\partial \Sigma$. See the Supplemental Material (SM) for details.

 Despite high nonlinearity of the action, the path integral can be performed explicitly since integrating out $\beta$'s localizes $\gamma$ to  holomorphic maps ($\bar{\partial} \gamma=\partial \bar{\gamma}=0$) \cite{Komatsu:2025sqo,Komatsu:2025dqv}. Using the path integral, one can also deduce the operator product expansion (OPE). Computing the OPE of the stress tensor of matter fields,
\begin{equation}
    T(z)=-
    \partial X^\mu \partial X_\mu (z)-\beta \partial \gamma(z)+2q\{ \gamma,z\}+T_\chi (z) \, ,
\end{equation}
 one can read off the central charge of $S_{\beta\gamma}$ to be $c_{\beta\gamma}=2+24q$ \cite{Komatsu:2025sqo}. Here $\{ \gamma,z\}$ is the Schwarzian derivative, $\{ \gamma,z\}=(\partial^{3}\gamma\partial\gamma-\frac{3}{2}(\partial^2\gamma)^2)/(\partial \gamma)^2$. Thus, conformal anomaly cancellation requires
\begin{equation}\label{eq:anomaly}
    q=1-\frac{d+c_\chi}{24} \, .
\end{equation}
As discussed in \cite{Komatsu:2025sqo,Komatsu:2025dqv}, for the CLD to be well-defined, $q$ needs to satisfy $q>0$, which translates to the condition $d+c_{\chi}<24$. In particular, requiring the internal CFT to be unitary $(c_{\chi}>0)$ leads to a bound $d<24$.

\begin{figure}[!t]
    \centering
   \begin{tikzpicture}[scale=0.8]
       \draw[ultra thick] (0,-2) -- (0,2)  (1.8,-2) -- (1.8,2)  (4.5,-2) -- (4.5,2) (5+1,-2) -- (5+1,2);

       \draw[ ->] (-2,0) -- (8,0);

       \draw[very thick, red, snake it ] (0,-0.7) -- (1.8,-0.7);

         \draw[very thick, violet, snake it ]  (1.8,0.8) -- (4.5,0.8);

        \draw[very thick, blue, snake it ]   (4.5,1.5) -- (6,1.5);

        \draw[very thick, teal, snake it ]  (0,-1.5) -- (6,-1.5);


\node at (8.5,0)  {\scalebox{1.3}{$\gamma^1$}}; 

\node at (0.9,-0.32)  {\scalebox{1}{$\boldsymbol{w_1}$}}; 

\node at (3.15,1.25)  {\scalebox{1}{$\boldsymbol{w_2}$}}; 

\node at (5.25,1.9)  {\scalebox{1}{$\boldsymbol{w_3}$}};

\node at (3,-1.1)  {\scalebox{1}{$\boldsymbol{w_4}$}};

\end{tikzpicture}
\caption{Open strings between D0-branes placed along $\gamma^1$-direction separated by $\pi w_a R$, where $w_a, \;a=1,\cdots,4$ are fractional winding numbers.}
\label{fig:D0}
\end{figure}
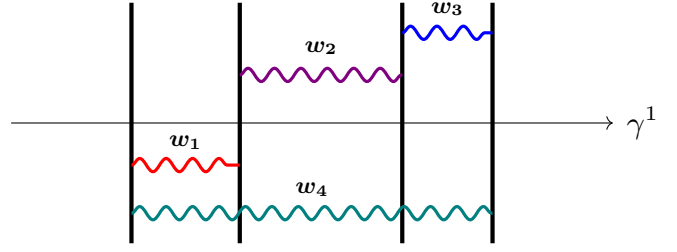

\noindent \textbf{Boundary condition.} To discuss amplitudes of open strings, we need to specify boundary conditions (bc) at $\partial \Sigma$. We choose the one corresponding to ``D0-branes" in the $\gamma$ plane i.e.~branes localized in $\gamma^{1}$ and extended in $\gamma^{2}$. See Fig.~\ref{fig:D0}. Consequently, we impose the Dirichlet bc for $\gamma^1$ and the Neumann bc for $\gamma^2$,
\begin{align}\label{eq:bc1}
\left. \gamma^1\right|_{\partial\Sigma} ={\rm const} \,,\qquad \left.\partial_n\gamma^2\right|_{\partial\Sigma}=0\,,
\end{align}
where $\partial_n$ is a derivative normal to the boundary $\partial\Sigma$. Taking the boundary to be ${\rm Im}\, z=0$ and using the holomorphicity of $\gamma$'s, \eqref{eq:bc1} can be rewritten as
\begin{align}\label{eq:bc2}
\left.\partial \gamma=-\bar{\partial}\bar{\gamma}\right|_{\partial\Sigma}\,.
\end{align}
Acting repeatedly the derivative parallel to the boundary $\partial+\bar{\partial}$ on \eqref{eq:bc2}, one can show $\left.\partial^{n}\gamma=-\bar{\partial}^{n}\bar{\gamma}\right|_{\partial \Sigma}$ for any positive integer $n$. This implies  $\left.(\partial-\bar{\partial})\varphi\right|_{\partial\Sigma}=\left.i\partial_y \varphi\right|_{\partial\Sigma}=0$, where we used the real coordinates $z=x+iy$. The bc for $\beta$'s can be determined by requiring the boundary contribution from the variation of $\gamma^{2}$ to vanish:  

\begin{equation}
    0= \frac{1}{2\pi} \int_{y=0} dx \left[ (\beta+\bar{\beta})+iq\, \partial_x \left( \frac{\partial_y\varphi}{\partial \gamma}\right) \right]\delta \gamma^2 \, .
\end{equation}
Since the second term in the square bracket vanishes due to $\partial_y \varphi=0$, we get the following bc for $\beta$'s:
\begin{align}
\left.\beta+\bar{\beta}\right|_{\partial\Sigma}=0\,.
\end{align}

\noindent\textbf{Vertex operators and four-point amplitudes.} We consider open-string vertex operators
\begin{equation}\label{vertex_operator}
      \mathcal{V}_{a}(x_a)=         e^{ik_a\cdot X(x_a)+iw_aR\int^{x_a} (\beta dz'  -\bar{\beta}d\bar{z}' )} \,,
 \end{equation}
These vertex operators create open strings stretched between two D0-branes separated by $\pi w_a R$ as can be seen in the OPE with $\gamma^1$:
\begin{align}
\gamma^1(z,\bar{z})\,  \mathcal{V}_a(x) \sim -\frac{iw_aR}{2} \log\left( \frac{z-x}{\bar{z}-x}\right) \mathcal{V}_a(x)
\end{align}
By computing correlation functions using a path integral, one can read off the conformal dimension of $\mathcal{V}_a$,  $h=\bar{h}=k_a^2+q$. Thus, the on-shell condition reads
\begin{align}
M_a^2=-k_a^2=q-1\,.
\end{align}

In the presence of vertex operators \eqref{vertex_operator}, the path integral of $\gamma(z)$ localizes to the ``Mandelstam map" \cite{Mandelstam:1985ww}, $\gamma(z)=R\rho(z)$ with
\begin{align}\label{eq:Mandelstam}
\rho (z)=\sum_{k}w_k  \log (z-x_k)\,.
\end{align}
As described in the Supplemental Material, this localization allows us to evaluate the worldsheet correlation function of matter fields as
\begin{align}
\begin{aligned}
&G_{4}^{\rm matter}=\mathcal{N}_4 \prod_{i<j}^{4}x_{ij}^{2k_i\cdot k_j}\delta_{\sum_{a}w_a}\delta^{d}\left(\sum_{i}k_i\right)\\
&\times\left[\frac{(w_1w_2w_3w_4)(w_1+w_3)^2}{(x_{12}x_{14}x_{23}x_{34})^{2}}(x_c-a_{+})(x_c-a_{-})\right]^{\frac{q}{2}}
\end{aligned}
\end{align}
where $x_c=(x_{12}x_{34})/(x_{13}x_{24})$ is the cross ratio and $a_{\pm}$ are given by
\begin{align}\label{eq:apmfirst}
a_{\pm}=-\frac{w_1w_4+w_2w_3 \pm 2\sqrt{w_1w_2w_3w_4}}{(w_1+w_3)^2} \, ,
\end{align}
while $\mathcal{N}_4$ is an overall factor independent of worldsheet and target-space kinematics. Combined with the correlation functions of $bc$-ghosts, this leads to the following expression for the four-point amplitudes
\begin{align}\label{eq:minfinal}
&\mathcal{A}_4 =\\ &\tilde{\mathcal{N}}_4\int_{0}^{1} dx \, x^{-s+q-2} (1-x)^{-t+q-2}  (x-a_{+})^{\frac{q}{2}}(x-a_{-})^{\frac{q}{2}} \, ,\nonumber
\end{align}
where $s=-(k_1+k_2)^2$ and $t=-(k_2+k_3)^2$ are Mandelstam variables and $\tilde{\mathcal{N}}_4$ is an overall factor independent of Mandelstam variables. Note that we used the PSL$(2,\mathbb{R})$ invariance to fix $x_{1,2,4}$ to $(0,1,\infty)$. 
Compared with \eqref{Mandelstam}, \eqref{eq:minfinal} has a few differences: 
\begin{enumerate}
\item  \eqref{eq:minfinal} is not $s,t$-crossing symmetric in general because of the extra factor $(x-a_{+})^{\frac{q}{2}}(x-a_{-})^{\frac{q}{2}}$.
\item The parameters $b$ and $\delta$ are both fixed by $q$ and are not independent in \eqref{eq:minfinal}:
\begin{align}\label{eq:relqpara}
\delta=\frac{q}{2}\,,\qquad b=1-q\,.
\end{align}
\end{enumerate}
The first issue can be resolved by tuning $w_a$'s. A particularly convenient choice is $w_1=w_3$, under which $a_{\pm}$ satisfy $a_{+}=1-a_{-}$ and are given by
\begin{align}
a_{\pm}=\frac{1}{2}\pm \frac{i}{2}\sqrt{r(2+r)}\,,
\end{align}
with $r=w_2/w_1$. With this choice, the parameter $\lambda$ in \eqref{Mandelstam} is given by
\begin{align}
\lambda=\frac{1}{(1+r)^2}\,.
\end{align}
Note that, for $q=0$, the amplitude reduces to the standard open bosonic amplitude as can be seen in \eqref{eq:relqpara}.

\noindent\textbf{Generalization by dressing.} The second issue can be overcome by making use of the internal CFT $S_{\chi}$. For concreteness, let us assume that $S_{\chi}$  contains an extra free boson $y(z,\bar{z})$ with an action
\begin{equation}\label{eq:freeboson}
     S_y=\int \frac{d^2z}{2\pi}\, \partial y \bar{\partial}y \, .
\end{equation}
We then dress the vertex operators $\mathcal{V}_{2,4}$ as
\begin{align}
\tilde{\mathcal{V}}_2=\mathcal{V}_2e^{i p y }\,,\quad \tilde{\mathcal{V}}_4=\mathcal{V}_4e^{-i p y }\,.
\end{align}
This modifies the on-shell condition to $M_{2,4}^2=-k_{2,4}^2=q+p^2-1$ and \eqref{eq:minfinal} to
\begin{align}
&\mathcal{A}_4 /\tilde{\mathcal{N}}_4=\\ &\int_{0}^{1} dx \, x^{-s+q+p^2-2} (1-x)^{-t+q+p^2-2}  (x-a_{+})^{\frac{q}{2}}(x-a_{-})^{\frac{q}{2}} \, ,\nonumber
\end{align}
changing \eqref{eq:relqpara} to
\begin{align}
\delta=\frac{q}{2}\,,\qquad b=1-q-p^2\,.
\end{align}
We can thus reproduce \eqref{Mandelstam} with general parameters from the worldsheet. In particular, by setting $p^2=1-q$, the lightest exchanged particle can be made massless as is the case with the original Veneziano amplitude.

\section{Extensions}
An advantages of having a worldsheet formulation is that it is straightforward to generalize the analysis beyond open-string four-point amplitudes. Here we present two generalizations: higher-point open string amplitudes and closed-string four-point amplitudes.
\subsection{Higher-point amplitudes}
Let us extend the computation to $n(> 4)$-point amplitudes. Using the vertex operators \eqref{vertex_operator} and
evaluating the path integral, we obtain
\begin{align}\label{npoint_correlator}
\begin{aligned}
  G^{\rm matter}_n =&\mathcal{N}_n \prod_{1\leq i< j\leq n} x_{ij}^{2 k_i\cdot k_j-q} \delta_{\sum_a w_a} \delta^d\left(\sum_i k_i\right)\\
  & \times \left|\Delta (P_{n})\prod_{k=1}^{n}w_k\right|^{\frac{q}{2}}   \,.
\end{aligned}  
\end{align}
Here $\Delta(P_{n})$ is a discriminant of a $(n-2)$ degree polynomial associated with the Mandelstam map $\rho(z)$ \eqref{eq:Mandelstam}
\begin{align}
\begin{aligned}
P_{n} (z)&\equiv\left(\prod_{k=1}^{n}(z-x_k)\right) \partial \rho(z)\\&=\sum_{k=1}^{n}w_k\prod_{j\neq k}(z-x_j)\,,
\end{aligned}
\end{align}
Written explicitly, it is given by
\begin{align}
\begin{aligned}
   & \Delta (P_{n})
= \left(\sum_{k=1}^{n}w_kx_k\right)^{2n-6} \prod_{I< J}^{n-2} (Z_I-Z_J)^2 \, ,
\end{aligned}
\end{align}
where $Z_I$'s are zeros of $\partial \rho(z)$, $\partial \rho (Z_I)=0$. As shown in the SM, $\Delta (P_n)$ transforms under PSL$(2,\mathbb{R})$, $x_k \to \frac{a x_k+b}{cx_k +d}$ $(ad-bc=1)$, as
\begin{align}
\Delta (P_n) \to \frac{\Delta (P_n)}{\prod_{k}(cx_k+d)^{2n-6}}\,,
\end{align}
guaranteeing that \eqref{npoint_correlator} behaves as a correlation function of (quasi-)primary operators of dimension $1$. 

As usual, scattering amplitudes can be computed by fixing three points using PSL$(2,\mathbb{R})$, including the ghost contribution and integrating over the remaining moduli. For instance, setting $x_{1,4,5}$ to $(0,1,\infty)$, the five-point amplitude reads
\begin{align}\label{eq:fivepoint}
&\frac{\mathcal{A}_5}{\tilde{\mathcal{N}}_5}=\int_{0\leq \chi_1<\chi_2\leq 1} \!\!\!\!\!\! \!\!\!\!\!\!\!\!\!\!\!d\chi_1 d\chi_2 \,\chi_1^{-s_{12}+q-2}(1-\chi_1)^{-s_{24}+q-2}\\
&\times\chi_2^{-s_{13}+q-2}(1-\chi_2)^{-s_{34}+q-2}(\chi_2-\chi_1)^{-s_{23}+q-2}\Delta(Q_5)^{\frac{q}{2}}\,, \nonumber
\end{align}
where $\tilde{\mathcal{N}}_5$ is an overall factor independent of the Mandelstam variables, $s_{ij}=-(k_i+k_j)^2$, and $\chi_{1,2}$ are two conformal cross ratios defined by, 
\begin{equation}
    \chi_1= \frac{ x_{12}x_{45} }{x_{14} x_{25} },\quad \chi_2= \frac{ x_{13}x_{45} }{x_{14} x_{35} }\,.
\end{equation}
$\Delta (Q_5)$ is a discriminant of a polynomial
\begin{align}
\begin{aligned}
Q_5(z)&=\left.\lim_{x_{1,4,5}\to (0,1,\infty)}\frac{P_5(z)}{x_5}\right|_{x_2 \to \chi_1,x_{3}\to \chi_2}\\
&=a_3 z^3+a_{2} z^2+a_{1}z+a_0 \, ,
\end{aligned}
\end{align}
with
\begin{align}
&a_3=w_5, \,\qquad a_0= w_1 \chi_1 \chi_2\,,\nonumber\\
&a_{2} = -\left( w_4+w_5+(w_2+w_5) \chi_1+(w_5+w_3) \chi_2 \right) \, ,\\
       &a_{1}= -\left((w_1+w_3) \chi_1+ (w_1+w_2) \chi_2+(w_1+w_4)\chi_1\chi_2 \right) \, .\nonumber
\end{align}

Analogous to the four-point case, the amplitude \eqref{eq:fivepoint} is not crossing symmetric in general but can be made partially crossing symmetric by tuning $w_i$'s. For instance, by setting $w_1=w_4$ and $w_2=w_3$, the amplitude becomes symmetric with respect to two independent exchanges, $k_{1}\leftrightarrow k_4$ and $k_{2}\leftrightarrow k_3$. One can also make the lightest exchanged particles in some of the channels massless by dressing some of the vertex operators by the free boson \eqref{eq:freeboson} (for instance, by $\tilde{\mathcal{V}}_2=\mathcal{V}_2e^{i p y }$, $ \tilde{\mathcal{V}}_3=\mathcal{V}_3e^{-i p y }$) and tuning the momentum $p$.

Importantly, for higher-point amplitudes, one cannot make them fully crossing-symmetric or render the lightest exchanged particles massless in every channel. This is because the winding conservation, $\sum_i w_i=0$, forces external particles into unequal roles; some $w_i$ must be positive while others are negative.
\subsection{Closed string amplitudes}
To compute closed-string amplitudes, we simply use \eqref{CLD} without the boundary term. As shown in \cite{Komatsu:2025sqo}, the resulting action is Weyl invariant. 

Inserting vertex operators in the bulk of the worldsheet, $\mathcal{V}_a=e^{ik_a\cdot X+iw_aR\int^{z_a } (\beta dz'  -\bar{\beta}d\bar{z}' )}$, dressing them by the free boson as $\tilde{\mathcal{V}}_2=\mathcal{V}_2e^{ip y}$ and $\tilde{\mathcal{V}}_4=\mathcal{V}_4e^{-ip y}$, and performing the path integral, we obtain
\begin{align}\label{closed_string_amplitude}
    &\frac{\mathcal{A}_{4}^{\text{closed}} }{\tilde{\mathcal{N}}_4^{\rm closed}}=\\
    &\int_{\mathbb{C}} d^2z\, |z|^{-\frac{s}{2}-2b-2} |1-z|^{-\frac{t}{2}-2b-2} |(z-a_{+})(z-a_{-})|^q \, ,\nonumber
\end{align}
with $b=1-q-\frac{p^2}{4}$ and $a_{\pm}$ given by \eqref{eq:apmfirst}.

Similarly to the open string case, \eqref{closed_string_amplitude} does not exhibit any crossing symmetry for general $w_i$'s but can be made $s$-$t$ symmetric by setting $w_1=w_3$;~i.e. symmetric with respect to $z\leftrightarrow 1-z$. However, it cannot be made $s$-$t$-$u$ symmetric; i.e.~$z\leftrightarrow 1-z \leftrightarrow \frac{z}{z-1}$. The reason for this limitation is again the winding conservation, which puts external particles on unequal footings.

Relatedly, although the complex integral \eqref{closed_string_amplitude} can be evaluated explicitly  using the tricks by Kawai-Lewellen-Tye \cite{Kawai:1985xq}, the result does not factorize simply into two open-string amplitudes \eqref{Mandelstam}. Instead, it is given by a sum of product of Appell's hypergeometric function $\mathbb{F}_1(a,b_1,b_2,c|x,y)$. See the SM for details.

It would be interesting to find a worldsheet action that gives rise to a generalized Virasoro-Shapiro amplitude with full crossing symmetry. To achieve it, one needs to relax the winding conservation, which forces the light-cone-gauge-like map. One possibility is to find a mechanism that localizes the worldsheet to a map with the ``Witten-type" interaction vertices \cite{Witten:1985cc}.

\section{Conclusion}
We presented a worldsheet action reproducing the generalized Veneziano amplitudes, originally introduced by Mandelstam. Using the same action, we also provided generalization to open-string higher point amplitudes and closed-string analogs. Being the first worldsheet description of generalized Veneziano amplitudes, our construction motivates further studies of the subject. To list a few,
\begin{itemize}
\item Unitarity of the generalized Veneziano amplitudes for $d=4$ was analyzed in \cite{Haring:2023zwu} by studying the positivity of their partial-wave coefficients. It would be useful to revisit this question from a worldsheet perspective, extending the result in \cite{Haring:2023zwu} to general dimensions and establishing no-ghost theorem from the worldsheet \footnote{Our construction requires $q>0$, i.e.~$d+c_{\chi}<24$. However, because of the unconventional structure of CLD, it is unclear if this condition is sufficient for the unitarity of spacetime amplitudes. To clarify this issue, one has to derive the no-ghost theorem for the CLD.}.
\item It would be interesting to look for the worldsheet actions reproducing other generalized Veneziano amplitudes, for instance the hypergeometric amplitudes discussed in \cite{Cheung:2023adk}.
\item Our construction gave closed string amplitudes with only partial crossing symmetry. As mentioned already, developing a worldsheet model that produces a fully crossing-symmetric generalized closed string amplitude would be highly illuminating.
\item The generalized Veneziano amplitude \eqref{Mandelstam} bears resemblance with off-shell string amplitudes computed from string field theory \cite{Rastelli:2007gg}. It would be interesting to understand possible connectons, if any.
\end{itemize}

\vspace{10pt}
\noindent 

\bibliographystyle{apsrev4-1}
\bibliography{ref}

\clearpage

\onecolumngrid
\appendix

\clearpage

\onecolumngrid
\appendix

{\begin{center}\bf \Large{Supplemental Material}\end{center} }

\section{Conformal invariance of our worldsheet action}
In this appendix, we find how our CLD worldsheet action with the boundary term \eqref{CLD} transforms under the Weyl transformation of the worldsheet metric: 
\begin{equation}
    g_{ab}\to g'_{ab}=e^{2w(\tau,\sigma)} g_{ab} \, .
\end{equation}
We first write down the covariant form of the CLD action: 
\begin{equation}\label{CLD_covariant}
   S_{\rm CLD}=\frac{2}{\pi}\left[ \frac{1}{4}\int_{\mathcal{M}}d^2\sigma \sqrt{-g} \,g^{ab} \partial_a \varphi \,\partial_b \varphi+\frac{1}{2} \int_{\mathcal{M}}d^2\sigma \sqrt{-g}  \,\hat{R}\varphi  +\int_{\partial \mathcal{M}}ds\, \hat{k}\, \varphi  \right]\, ,
\end{equation}
where 
\begin{equation}\label{covariant_phi}
    \varphi = \log\Big[ \left(\frac{1}{4}g^{ab} \partial_a \gamma \,\partial_b \bar{\gamma} +\frac{i}{4} \epsilon^{cd} \partial_c \gamma \, \partial_d \bar{\gamma}\right)/R^2 \Big] \, .
\end{equation}
An overall factor of $q/2$ has been suppressed and will be reinstated at the end. We note down the variations of various geometric quantities under the Weyl transformation, that will be useful for our computations.  

\vspace{0.5cm}

\begin{itemize}
    \item Variation of metric determinant: 
    \begin{equation}
    \sqrt{-g'}= e^{2w}\sqrt{-g} \, .
\end{equation}

\item Variation of Christoffel connection: 
\begin{equation}
    \begin{split}
        (\Gamma')^a_{bc}&=\frac{1}{2}(g')^{ad}\Big[ \partial_b g'_{cd}+\partial_c g'_{bd}-\partial_d g'_{bc}\Big]\\
        &=\frac{1}{2}g^{ad}\Big[ \partial_b g_{cd}+\partial_c g_{bd}-\partial_d g_{bc}\Big] +\frac{1}{2}g^{ad} \cdot 2\left( g_{cd}\partial_b w +g_{bd}\partial_c w -g_{bc}\partial_d w\right)\\
        &= \Gamma^a_{bc}+ \left( \delta^a_c \,\partial_b w+\delta^a_b \, \partial_c w-g^{ad}g_{bc}\,\partial_d w \right) \, .
    \end{split}
\end{equation}

\item Variation of $\sqrt{-g} \hat{R}$:
\begin{equation}
    \sqrt{-g'} \hat{R}'=\sqrt{-g} \left( \hat{R}-2\nabla_a \partial^a w\right) \, .
\end{equation}

\item Variations of unit vectors at $\partial \mathcal{M}$: We denote by $t^a$ the unit vector tangent to the boundary of the worldsheet $\partial \mathcal{M}$, and by $n^a$ the outward-pointing unit vector normal (perpendicular) to $\partial \mathcal{M}$. For spacelike or timelike such unit vectors, we thus obtain,
\begin{equation}
    g_{ab}\,t^at^b=\pm 1,\quad (g')_{ab}\, (t')^a (t')^b=\pm 1 \Rightarrow (t')^a=e^{-w}\, t^a \, .
\end{equation}
Similarly,
\begin{equation}
    (n')^a=e^{-w}\, n^a \,.
\end{equation}

\item Variation of the geodesic curvature: The geodesic curvature is defined by: 
\begin{equation}
    \hat{k}=\pm t^a \,n_b \nabla_a t^b \, ,
\end{equation}
with upper and lower signs are for Lorentzian and Euclidean worldsheets respectively. Under Weyl transformation
\begin{equation}
    \begin{split}
        \hat{k}' &= \pm (t')^a (n')_b \nabla'_a(t')^b\\
        &= \pm e^{-w} t^a \cdot e^w n_b\cdot \left[\partial_a(e^{-w}t^b)+(\Gamma')^{b}_{ac} \,e^{-w} t^c \right] \\
        &= \pm e^{-w} t^a n_b (\partial_a t^b +\Gamma^b_{ac} t^c) \pm e^{-w} \Big[ -t^a \textcolor{blue}{n_b t^b} \partial_a w+ \textcolor{blue}{t^an_a} \partial_c w\, t^c +t^a \textcolor{blue}{n_c} \partial_a w \, \textcolor{blue}{t^c} - n_b \partial_d w\, g^{bd}\, g_{ac}t^a t^c \Big]\\
        &= e^{-w} k\mp e^{-w} n^d \partial_d w\, ||t||^2\\
        &= e^{-w} \left( \hat{k}+n^d\partial_d w\right) \, ,
    \end{split}
\end{equation}
where we have used $\textcolor{blue}{t^a n_a=0}$, and $g_{ac}t^a t^c=||t||^2=-1$ when the boundary is timelike, and  $||t||^2=+1$ when it is spacelike so that $\mp\mp=+$.

\item Variation of line element at $\partial \mathcal{M}$: Writing explicitly,
\begin{equation}
    ds=\sqrt{-g_{00}} \, d\tau \quad \text{(for timelike boundary)}, \quad  ds=\sqrt{g_{11}} \, d\sigma \quad \text{(for spacelike boundary)}\, ,
\end{equation}
And so, under the Weyl transformation,
\begin{equation}
    ds'=e^{w} ds \, .
\end{equation}

\end{itemize}

With these results, we now compute variations of different terms of our CLD action under infinitesimal version of the Weyl transformation: $  g'_{ab}=e^{2\delta w} g_{ab}\approx (1+2\delta w) \, g_{ab}$. We first observe that since both $g^{ab}$ and $\epsilon^{ab}$ changes in the same way: $g^{ab} \to (1-2\delta w) g^{ab},\; \epsilon^{ab} \to (1-2\delta w) \epsilon^{ab}$, the $\varphi$ will be shifted as
\begin{equation}
   \varphi\to \varphi'= (\varphi-2\delta w) \, .
\end{equation}
With this,
\vspace{0.5cm}
\begin{itemize}
    \item Variation of $\hat{R}\varphi$ term:
\begin{equation}
    \begin{split}
         \int_{\mathcal{M}}d^2\sigma \sqrt{-g'}  \,\hat{R'}\varphi'  &=  \int_{\mathcal{M}}d^2\sigma \sqrt{-g} \,e^{2\delta w} \,e^{-2\delta w}(\hat{R}-2\nabla_a\partial^a \delta w)\,(\varphi -2\delta w)\\
         &=\int_{\mathcal{M}}d^2\sigma \sqrt{-g} \Big( \hat{R}\varphi -2\hat{R} \,\delta w-2\varphi \nabla_a\partial^a\delta w\Big) +\mathcal{O} (\delta w^2) \, .
    \end{split}
\end{equation}
For the last term within bracket in r.h.s, we use integration by parts:
\begin{equation}
    \begin{split}
        \int_{\mathcal{M}}d^2\sigma \sqrt{-g} \,\varphi \nabla_a\partial^a\delta w& = \int_{\mathcal{M}}d^2\sigma \sqrt{-g} \,[ \nabla_a(\varphi \partial^a \delta w)-\partial_a \varphi \partial^a \delta w]\\
        &= \int_{\partial \mathcal{M}}ds\, n_a \varphi\, \partial^a \delta w - \int_{\mathcal{M}}d^2\sigma \sqrt{-g} \, [\nabla^a(\delta w\,\partial_a\varphi) -\delta w\,\nabla^a\partial_a\varphi ]\\
        &= \int_{\partial \mathcal{M}}ds\, n_a \varphi\, \partial^a \delta w -  \int_{\partial \mathcal{M}}ds \, n^a \delta w \, \partial_a\varphi+ \int_{\mathcal{M}}d^2\sigma \sqrt{-g} \, (\nabla^a\partial_a\varphi)\,\delta w \, .
    \end{split}
\end{equation}
Hence,
\begin{align}
    \begin{aligned}
        &\delta\left(\int_{\mathcal{M}}d^2\sigma \sqrt{-g}  \,\hat{R}\varphi  \right) =\\     &-2\int_{\mathcal{M}}d^2\sigma \sqrt{-g}  \,\hat{R}\,\delta w 
        -2\int_{\partial \mathcal{M}}ds\, n_a \varphi\, \partial^a \delta w +2 \int_{\partial \mathcal{M}}ds \, n^a \delta w \, \partial_a\varphi  -2\int_{\mathcal{M}}d^2\sigma \sqrt{-g} \, (\nabla^a\partial_a\varphi)\,\delta w \, .
    \end{aligned}
\end{align}

\item  Variation of $\hat{k}\varphi$ term:
    \begin{equation}
        \begin{split}
            \int_{\partial \mathcal{M}} ds' \hat{k}'\varphi' &=  \int_{\partial \mathcal{M}} ds e^{\delta w} \cdot e^{-\delta w} (\hat{k}+n^a\partial_a \delta w) (\varphi-2\delta w)\\
            &= \int_{\partial \mathcal{M}} ds \,\hat{k}\,\varphi -2\int_{\partial \mathcal{M}} ds \, \hat{k} \, \delta w +\int_{\partial \mathcal{M}} ds \,\varphi\, n^a \partial_a \delta w \, .
        \end{split}
    \end{equation}
Thus, 
\begin{equation}
    \delta \left(\int_{\partial \mathcal{M}} ds \,\hat{k}\,\varphi \right) = -2\int_{\partial \mathcal{M}} ds \, \hat{k} \, \delta w +\int_{\partial \mathcal{M}} ds \,\varphi\, n^a \partial_a \delta w  \, .
\end{equation}

\item Variation of the kinetic term $(\partial \varphi)^2$: 
\begin{equation}
    \begin{split}
       \delta\left( \int_{\mathcal{M}}d^2\sigma \sqrt{-g'}\, (g')^{ab} \partial_a \varphi \,\partial_b \varphi\right) &=  -2\int_{\mathcal{M}}d^2\sigma \sqrt{-g} \, g^{ab} \cdot2\partial_a (\delta w) \,\partial_b \varphi\\
        &= -4 \int_{\mathcal{M}}d^2\sigma \sqrt{-g} \, [\nabla^a(\delta w \partial_a \varphi) -\delta w\nabla^a\partial_a\varphi]\\
        &=-4\int_{\partial \mathcal{M}}ds \,n^a\delta w \,\partial_a\varphi +4 \int_{\mathcal{M}}d^2\sigma \sqrt{-g} \, \delta w\nabla_a(\partial^a \varphi) \, .
    \end{split}
\end{equation}
\end{itemize}

Adding them, we obtain
\begin{equation}\label{eq:Weylfinal}
    \delta \Big(S_{\rm CLD}\Big)= \frac{2}{\pi} \left[ -\int_{\mathcal{M}}d^2\sigma \sqrt{-g}  \,\hat{R}\,\delta w -2\int_{\partial\mathcal{M}} ds \,\hat{k}\,\delta w \right]
\end{equation}
In particular all the other terms: $\int_{\partial \mathcal{M}} ds \,\varphi\, n^a \partial_a \delta w$, $\int_{\partial \mathcal{M}}ds \,n^a \partial_a\varphi \,\delta w$, and $\int_{\mathcal{M}}d^2\sigma \sqrt{-g} \, \nabla_a(\partial^a \varphi) \, \delta w$ cancel out between different terms of the CLD action once their relative coefficients are taken into account. The expression \eqref{eq:Weylfinal} reproduces the expected structure of the Weyl anomaly in the presence of a boundary. By reinstating the overall $q/2$ factor, we correctly reproduce the central charge from the CLD action, $c_{\rm CLD}=24q$ \cite{Komatsu:2025sqo}.

\section{Mandelstam formula: derivation and conformal property}

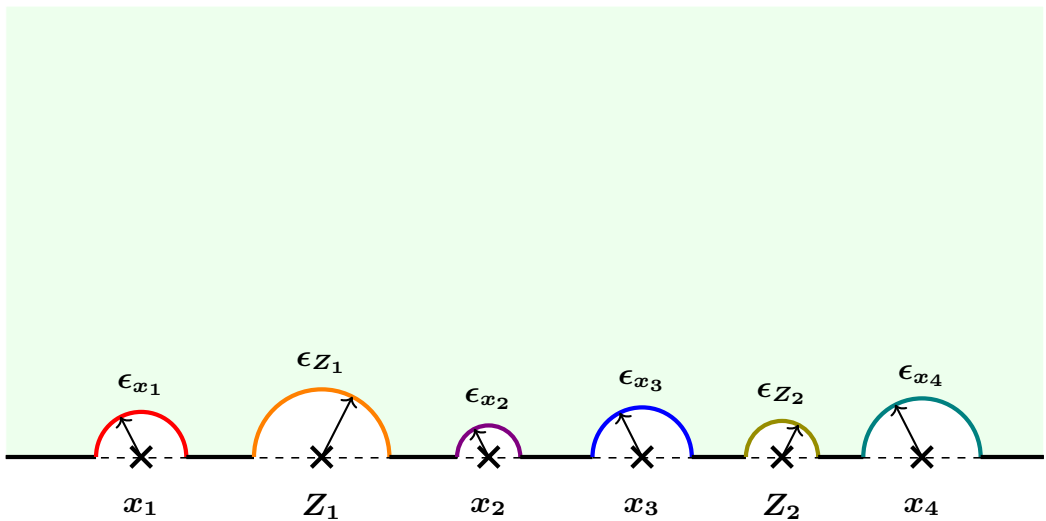
\begin{figure}[!htb]
    \centering
\begin{tikzpicture} [scale = 0.6]


\draw[white, fill=green!7!]  (0,0) -- (23,0) -- (23,10) -- (0,10) -- (0,0);

\draw[very thick, dashed] (0,0) -- (23,0);


\draw[ultra thick] (0,0) -- (2,0);

\draw[red, ultra thick, fill=white] (2,0) arc (180:0:1);

\draw[ultra thick] (4,0) -- (5.5,0);

\draw[orange, ultra thick, fill=white] (5.5,0) arc (180:0:1.5);

\draw[ultra thick] (8.5,0) -- (10,0);

\draw[violet, ultra thick, fill=white] (10,0) arc (180:0:0.7);

\draw[ultra thick] (11.4,0) -- (13,0);

\draw[blue, ultra thick, fill=white] (13,0) arc (180:0:1.1);

\draw[ultra thick] (15.2,0) -- (16.4,0);

\draw[olive, ultra thick, fill=white] (16.4,0) arc (180:0:0.8);

\draw[ultra thick] (18,0) -- (19,0);

\draw[teal, ultra thick, fill=white] (19,0) arc (180:0:1.3);

\draw[ultra thick] (21.6,0) -- (23,0);

\draw (3,0) node[cross] {};

\draw (7,0) node[cross] {};

\draw (10.7,0) node[cross] {};

\draw (14.1,0) node[cross] {};

\draw (17.2,0) node[cross] {};

\draw (20.3,0) node[cross] {};


\node at (3,-1.1)  {\scalebox{1.3}{$\boldsymbol{x_1}$}}; 

\node at (7,-1.1)  {\scalebox{1.3}{$\boldsymbol{Z_1}$}}; 

\node at (10.7,-1.1)  {\scalebox{1.3}{$\boldsymbol{x_2}$}}; 

\node at (14.1,-1.1)  {\scalebox{1.3}{$\boldsymbol{x_3}$}}; 

\node at (17.2,-1.1)  {\scalebox{1.3}{$\boldsymbol{Z_2}$}}; 

\node at (20.3,-1.1)  {\scalebox{1.3}{$\boldsymbol{x_4}$}};

\node at (3,1.6)  {\scalebox{1.3}{$\boldsymbol{\epsilon_{x_1}}$}}; 

\node at (7,2.1)  {\scalebox{1.3}{$\boldsymbol{\epsilon_{Z_1}}$}}; 

\node at (10.7,1.3)  {\scalebox{1.3}{$\boldsymbol{\epsilon_{x_2}}$}}; 

\node at (14.1,1.7)  {\scalebox{1.3}{$\boldsymbol{\epsilon_{x_3}}$}}; 

\node at (17.2,1.4)  {\scalebox{1.3}{$\boldsymbol{\epsilon_{Z_2}}$}}; 

\node at (20.3,1.8)  {\scalebox{1.3}{$\boldsymbol{\epsilon_{x_4}}$}};


\draw[thick, ->] (3,0) -- (2.54,0.89);

\draw[ thick, ->] (7,0) -- (7.68, 1.336);  

\draw[ thick, ->] (10.7,0) -- (10.3822, 0.624);  

\draw[ thick, ->] (14.1,0) -- (13.606, 0.98);  

\draw[ thick, ->] (17.2,0) -- (17.563, 0.713);  

\draw[ thick, ->] (20.3,0) -- (19.709, 1.158);

\end{tikzpicture}

\caption{To regularise CLD action, we excise semi-circular discs from the upper half plane around insertion points $\{ x_k\}_{k=1}^{n}$, interaction points $\{Z_I\}_{I=1}^{n-2}$, and the point at infinity $x_\infty$. }
\end{figure}
In this appendix, we evaluate the chiral composite dilaton action $\Gamma[\gamma,\bar{\gamma}]$ in \eqref{CLD_covariant} when the fields $(\gamma,\bar{\gamma})$ are localized at the Mandelstam maps $(\gamma,\bar{\gamma})=R(\rho,\bar{\rho})$:
\begin{equation}
    \rho(z)= \sum_{k=1}^{n}w_k\log(z-x_k),\qquad   \bar{\rho}(\bar{z})= \sum_{k=1}^{n}w_k\log(\bar{z}-x_k)\,,
\end{equation}
and discuss its properties.

\subsection{Derivation of the Mandelstam formula.}
In what follows, we will work in a parameter regime of $\{w_k\}_{k=1}^{n}$ in which the interaction points $Z_I,\, I=1,\cdots,n-2$ are real. The final result thus obtained is analytic in $\{w_k\}$, and therefore can be analytically continued to arbitrary parameter regimes. 

Note that $\Gamma[\rho,\bar{\rho} ]$ can be viewed as the Liouville action associated with the conformal map from the worldsheet to the target space $(\rho,\bar{\rho})$. Its explicit evaluation has been discussed in detail in \cite{Mandelstam:1985ww}. We briefly review this derivation below, paying particular attention to the regularization subtleties involved.

To regularise $\Gamma[\rho,\bar{\rho}]$, we excise semi-circular discs from the upper half plane around the vertex operator insertion points $\{x_k\}_{k=1}^{n}$, the interaction points $\{Z_I\}_{I=1}^{n-2}$, and the point at infinity $x_\infty$. 

In passing, we rewrite $\partial \rho$ and $\bar{\partial}\bar{\rho}$ as:
\begin{equation}\label{rho-compact-form}
    \partial \rho (z) =\left(\sum_{k=1}^{n}w_k x_k\right)\frac{\prod_{I=1}^{n-2}(z-Z_I)}{\prod_{k=1}^{n}(z-x_k)},\quad \bar{\partial} \bar{\rho} (\bar{z}) =\left(\sum_{k=1}^{n}w_k x_k\right)\frac{\prod_{I=1}^{n-2}(\bar{z}-Z_I)}{\prod_{k=1}^{n}(\bar{z}-x_k)}\, ,
\end{equation}
where $Z_I$'s are zeros of $\partial \rho$.

\begin{itemize}
    \item \textbf{Kinetic part of CLD action:} We first rewrite the kinetic term as a boundary integral:
\begin{equation}
\begin{split}
\left(\Gamma[\rho]\right)_{\text{kinetic}}&=\frac{q}{2} \int_{\mathcal{M}} \frac{d^2x}{2\pi} \, \partial_{a}\varphi\,\partial^{a}\varphi=-\frac{q}{4\pi} \int_{\mathcal{M}} d^2x\, \varphi\, \partial_{a}\partial^{a}\varphi+ \frac{q}{4\pi} \int_{\partial \mathcal{M}} ds\,\varphi\, \partial_n \varphi \, ,
\end{split}
\end{equation} 
where $z=x^1+ix^2$ and $\bar{z}=x^1-ix^2$. Here $\partial_n$ is the normal derivative to $\partial\mathcal{M}$ in the outward direction. Since $\partial_a\partial^a\varphi=0$ in the bulk of the worldsheet, the only contribution to the kinetic part comes from the boundary integrals along semi-circular boundaries of the excised discs and the line segments between them:
\begin{equation}\label{kinetic}
	\left(\Gamma[\rho]\right)_{\text{kinetic}}= \frac{q}{4\pi} \int_{\partial \mathcal{M}} ds\,\varphi\, \partial_n \varphi =\frac{q}{4\pi} \left(\int_{\cup_{I} \partial D_{Z_I}} +\int_{\cup_k \partial D_{x_k}}+\int_{\partial D_{x_\infty}} +\int_{\text{line segment}}\right)ds\,\varphi\, \partial_n \varphi \, ,
\end{equation}
The normal derivatives are $ \partial_n=-\frac{1}{|z|}(z\,\partial+\bar{z}\,\bar{\partial})$ along the semi-circular disc boundaries, and $\partial_n=-\partial_2=-i(\partial -\bar{\partial})$ along the line segments on the real axis. From \eqref{rho-compact-form}, one can easily compute:
\begin{equation}
    \partial_n\varphi|_{z=\bar{z}}= -i\left( \partial \log [\partial\rho(z)]- \bar{\partial} \log[ \bar{\partial} \bar{\rho}(z)] \right)=0 \, ,
\end{equation}
where it is important that we have chosen $Z_I$'s as real. 

The contributions from the semi-circular discs are precisely $1/2$ of what is computed for closed string in \cite{Komatsu:2025dqv}. So we write it explicitly:
\begin{equation}
  \frac{1}{q}  \left(\Gamma[\rho]\right)_{\text{kinetic}} = \log \left[\left|\sum_{k=1}^{n}w_k x_k\right|^{-2} \delta_{\infty}^{-4}\,\prod_{I=1}^{n-2}\left|\partial^2\rho(Z_I)\right|^{-1/2}(2r_I)^{-1/2}\prod_{k=1}^{n}\left|w_k\right|\,\epsilon_{x_k}^{-1}\right] \, ,
\end{equation}
where $\{ \epsilon_{x_k}\}_{k=1}^{n}$ and $\{ r_I \}_{I=1}^{n-2}$ are radii of discs excised around $\{x_k\}_{k=1}^{n}$ on the worldsheet, and around $(\rho(Z_I),\bar{\rho} (\bar{Z_I}))$ on the target space respectively \cite{Komatsu:2025dqv}. We also have cut off $x$-space at $x=-\frac{1}{\delta_\infty}$ with $\delta_\infty$ being a small real number.

\item \textbf{Curvature part of CLD action:} For the curvature term of the CLD action:
\begin{equation}
    \left( \Gamma[\rho]\right)_{\text{curvature}}=\frac{q}{2\pi}\int_{\mathcal{M}} d^2z \left( \sqrt{-g} \,R\right)_{z,\bar{z}} \,\varphi (z,\bar{z}) +\frac{q}{\pi}\int_{\partial \mathcal{M}} ds \,k \,\varphi
\end{equation}
Similar to \cite{Komatsu:2025dqv}, the curvature on the worldsheet is concentrated along the semi-circular boundaries of the excised discs around the insertion points $\{ x_k\}_{k=1}^{n}$, and the point at infinity $x_\infty$. Explicitly,
\begin{equation}
    \left( \sqrt{-g}\, R \right)_{(z,\bar{z})} = 4\delta_\infty \cdot \delta\left( |z|-\frac{1}{\delta_\infty}\right) -2\sum_{k=1}^{n} \frac{1}{\epsilon_{x_k}} \delta\left(|z-x_k|-\epsilon_{x_k} \right) \, .
\end{equation}
Using this, we obtain:
\begin{equation}
    \frac{1}{q} \left( \Gamma[\rho]\right)_{\text{curvature}}= \log \left[ \left|\sum_{k=1}^{n}w_k x_k\right|^4 \,\delta_{\infty}^{8} \prod_{k=1}^{n} |w_k|^{-2} \epsilon_{x_k}^{2}  \right]\, .
\end{equation}

\item \textbf{Total contribution:}
Adding both kinetic and curvature parts of the CLD action, we get
\begin{equation}
    \frac{1}{q} \Gamma [\rho] = \log\left[ \prod_{k=1}^{n}|w_k|^{-1} \left|\sum_{k=1}^{n}w_k x_k\right|^2  \prod_{I=1}^{n-2}\left|\partial^2\rho(Z_I)\right|^{-1/2}  \frac{ \delta_\infty^4 \prod_{k=1}^{n}\epsilon_{x_k} }{ \prod_{I=1}^{n-2}(2r_I)^{1/2} } \right] \, .
\end{equation}
As in the closed-string case \cite{Komatsu:2025sqo,Komatsu:2025dqv}, we take $r_I=r,\, \forall I$, and then the on-shell CLD action contributes to the worldsheet correlator as:
\begin{equation}
    e^{-(\Gamma[\rho]-\Gamma_0)}= \prod_{k=1}^{n}|w_k|^{q} \left|\sum_{k=1}^{n}w_k x_k\right|^{-2q} \prod_{I=1}^{n-2}\left|\partial^2\rho(Z_I)\right|^{q/2}   \cdot (2r)^{q(n-2)/2} \prod_{k=1}^{n} \epsilon_{x_k}^{-q}\, ,
\end{equation}
where we divided the result by the disk partition function $e^{-\Gamma_0}$, i.e.~the exponentiated on-shell CLD action without any vertex operator insertions. 
\end{itemize}
The worldsheet correlation function is accompanied by the open string coupling constant $g_o^{n-2}$. Using this fact, we can get rid of the target space regularization parameter $r$ by renormalizing $g_s$:
\begin{equation}
    \boldsymbol{g}_o=(2r)^{q/2} g_o\, ,
\end{equation}
and we keep $\boldsymbol{g}_o$ as finite as physical open string coupling. We can also absorb $\prod_{k=1}^{n} \epsilon_{x_k}^{-q}$ in the renormalization of worldsheet vertex operators. After this, the renormalized on-shell CLD action is: 
\begin{equation}\label{Mandelstam_final}
      e^{-\Gamma^{\rm ren}[\rho]}= \prod_{k=1}^{n}|w_k|^{q} \left|\sum_{k=1}^{n}w_k x_k\right|^{-2q} \prod_{I=1}^{n-2}\left|\partial^2\rho(Z_I)\right|^{q/2}  \, .
\end{equation}
To make contact withe expression in the main text, we use \eqref{rho-compact-form} to get
\begin{align}
\prod_{I=1}^{n-2}\partial^2 \rho(Z_I)=\left(\sum_{k=1}^{n}w_k x_k\right)^{n-2}\frac{\prod_{I\neq J}(Z_I-Z_J)}{\prod_{I,k}(Z_I-x_k)}\,.
\end{align}
We can then rewrite \eqref{Mandelstam_final} as
\begin{align}
e^{-\Gamma^{\rm ren}[\rho]}=\left|\Delta (P_{n})\prod_{k=1}^{n}w_k\right|^{\frac{q}{2}} \prod_{1\leq i<j\leq n}|x_i-x_j|^{-q} \,,
\end{align}
where $\Delta (P_n)$ is a discriminant defined in the main text
\begin{align}
\Delta (P_{n})
= \left(\sum_{k=1}^{n}w_kx_k\right)^{2n-6} \prod_{I< J}^{n-2} (Z_I-Z_J)^2 \, .
\end{align}
\subsection{Conformal property}
We now study how $e^{-\Gamma^{\rm ren}[\rho]}$ transforms under the PSL($2,\mathbb{R}$) transformation $z \to (az+b)/(cz+d)$ with $ad-bc=1$. For this purpose, it is useful to recall that $(x_1-x_2)$ transforms under the PSL($2,\mathbb{R}$), $x_{1,2} \to (ax_{1,2}+b)/(cx_{1,2}+d)$, as
\begin{align}\label{eq:transbasic}
(x_1-x_2)\quad \to \quad \frac{x_1-x_2}{(cx_1+d)(cx_2+d)}\,. 
\end{align}
Another useful equality is
\begin{align}\label{eq:wjzj}
\sum_{j=1}^{n}w_jx_j=w_k \frac{\prod_{j(\neq k)}(x_k-x_j)}{\prod_{I}(x_k-Z_{I})}\,,
\end{align}
which one can derive by evaluating the residue of $\partial \rho(z)$ at $z=x_k$ using \eqref{rho-compact-form}:
\begin{align}
w_k=\Res_{z=x_k}\partial \rho(z)= \left(\sum_{j=1}^{n}w_j x_j\right)\frac{\prod_{I=1}^{n-2}(x_k-Z_I)}{\prod_{j(\neq k)}(x_k-x_j)}\,.
\end{align}
Using \eqref{eq:wjzj}, one can determine the PSL(2,$\mathbb{R}$) transformation of $\sum_k w_k x_k$ as
\begin{align}\label{eq:sumtrans}
\sum_k w_k x_k \quad \to \quad \frac{\prod_I (cZ_I +d)}{\prod_{k}(cx_k+d)}\left(\sum_k w_k x_k\right)\,.
\end{align}
From \eqref{eq:transbasic} and \eqref{eq:sumtrans}, we can derive the transformation rule of $\Delta (P_n)$,
\begin{align}
\Delta (P_n) \quad \to \quad \frac{\Delta (P_n)}{\prod_{k}(cx_k+d)^{2n-6}}\,.
\end{align}
\section{Computing the closed string amplitude using KLT-like factorization}
In this appendix, we compute the closed string amplitude \eqref{closed_string_amplitude} using a Kawai-Lwewllen-Tye-like factorization. Throughout this appendix, we assume the $s$-$t$ symmetric choice $a_{+}+a_{-}=1$. 

Writing $z=x+i\tilde{y}$ and $\bar{z}=x-i\tilde{y}$, the integral $\mathcal{I}$ in \eqref{closed_string_amplitude} takes the form
\begin{equation}
    \begin{split}
          \mathcal{I} = 2\int_{-\infty}^{\infty} dx \int_{-\infty}^{\infty} d\tilde{y} \, &(x+i\tilde{y})^{- s/4-b-1}  (1-x-i\tilde{y})^{-t/4-b-1} (x-a_{+}+i\tilde{y})^{q/2} (x-a_{-}+i\tilde{y})^{q/2}\\
          \times& (x-i\tilde{y})^{- s/4-b-1}  (1-x+i\tilde{y})^{-t/4-b-1} (x-a_{+}-i\tilde{y})^{q/2} (x-a_{-}-i\tilde{y})^{q/2} \, .
    \end{split}
\end{equation}
We note that the $\tilde{y}$-integral has branch points along the imaginary axis at $ \tilde{y} = \pm i x,\, \pm i(1-x), \, \pm i(x-a_+), \, \pm i (x-a_-)\, $. In what follows, we work in a parameter regime of $w_i$'s in which $a_{\pm}$ are real. Then we can safely rotate $\tilde{y}$-contour anti-clockwise by an angle $(\pi/2-\epsilon)$ without crossing any branch point. Along the deformed contour,  we have the change of variable: $\tilde{y}=e^{i\left( \frac{\pi}{2}-\epsilon\right)} y\simeq (iy+\epsilon y)$ with $\epsilon\to0_+ \,$. After this deformation, the integral becomes 
\begin{equation}
    \mathcal{I} = i\,\mathcal{I}_\zeta\, \mathcal{I}_\xi \, ,
\end{equation}
where
\begin{equation}
    \begin{split}
    \mathcal{I}_\zeta & =  \int d\zeta \,  (\zeta+i\delta)^{- s/4-b-1} (1-\zeta-i\delta)^{-t/4-b-1} (\zeta -a_{+}+i\delta)^{q/2} (\zeta-a_{-}+i\delta)^{q/2}  \, ,\\
      \mathcal{I}_\xi & =  \int d\xi  \, (\xi-i\delta)^{- s/4-b-1} (1-\xi+i\delta)^{-t/4-b-1}  (\xi-a_{+}-i\delta)^{q/2} (\xi-a_{-}-i\delta)^{q/2} \, ,
  \end{split}
\end{equation}
and $\xi=x+y$ and $\zeta=x-y$ are light-cone variables on the worldsheet, and we have defined $\delta=\epsilon y=\frac{1}{2}\epsilon (\xi-\zeta)$. Although the expressions appear to decouple, the branch structures of the two integrals remain correlated through the dependence on $\delta=\epsilon(\xi-\zeta)/2$ To make this explicit, we list the branch points of the $\xi$-integrand:  
\begin{equation}
    \begin{split}
        \xi_1 &=i\delta =\frac{i}{2}\epsilon (\xi-\zeta) =-\frac{i}{2}\epsilon \,\zeta +\mathcal{O}(\epsilon^2) \, ,\\
        \xi_2 &= 1+i\delta =1+\frac{i}{2}\epsilon(\xi-\zeta) =1+\frac{i}{2}\epsilon\,(1-\zeta) +\mathcal{O}(\epsilon^2) \, , \\
        \xi_3 &=a_{+}+i\delta = a_{+}+\frac{i}{2}\epsilon (\xi-\zeta) =a_{+}+\frac{i}{2} \epsilon\,(a_{+}-\zeta)+\mathcal{O}(\epsilon^2) \, , \\
        \xi_4 &= a_{-}+i\delta =a_{-}+\frac{i}{2} \epsilon(\xi-\zeta)=a_{-}+\frac{i}{2}\epsilon\,(a_{-}-\zeta) +\mathcal{O}(\epsilon^2) \, .
    \end{split}
\end{equation}
Without loss of generality we restrict to the ordering
\begin{equation}
    0<a_{+}<a_{-}<1 \, ,
\end{equation}
and evaluate  $\mathcal{I}_\zeta $ piecewise in different $\zeta$-intervals. For each case, the corresponding branch structure of $\mathcal{I}_\xi$ is different, and we compute it by different contour deformations. We then express our integrals in terms of Appell's hypergeometric function $\mathds{F}_1$, defined by
\begin{equation}
    \mathds{F}_1(a,b_1,b_2,c|x,y)=\frac{1}{\mathds{B}(a,c-a)}\int_0^1 dt \, t^{a-1}(1-t)^{c-a-1} (1-xt)^{-b_1} (1-yt)^{-b_2} \, ,
\end{equation}
where $\mathds{B}(X,Y)$ is the beta function
\begin{align}
\mathds{B}(X,Y)=\frac{\Gamma(X)\Gamma(Y)}{\Gamma(X+Y)}\,.
\end{align}

\subsection{$-\infty<\zeta<0$}
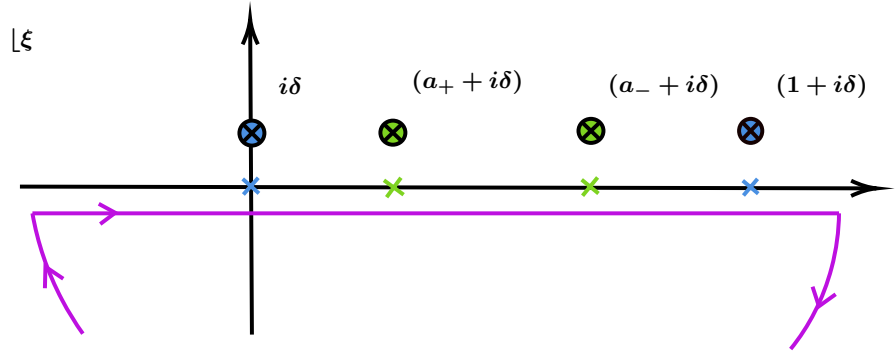
\begin{figure}[!htb]
    \centering
    \begin{tikzpicture}[x=0.75pt,y=0.75pt,yscale=-1,xscale=1, scale=0.9]

\draw [line width=1.5]    (121,220.5) -- (598,221.49) ;
\draw [shift={(601,221.5)}, rotate = 180.12] [color={rgb, 255:red, 0; green, 0; blue, 0 }  ][line width=1.5]    (14.21,-4.28) .. controls (9.04,-1.82) and (4.3,-0.39) .. (0,0) .. controls (4.3,0.39) and (9.04,1.82) .. (14.21,4.28)   ;
\draw [line width=1.5]    (251,303.5) -- (250.02,129.5) ;
\draw [shift={(250,126.5)}, rotate = 89.68] [color={rgb, 255:red, 0; green, 0; blue, 0 }  ][line width=1.5]    (14.21,-4.28) .. controls (9.04,-1.82) and (4.3,-0.39) .. (0,0) .. controls (4.3,0.39) and (9.04,1.82) .. (14.21,4.28)   ;
\draw  [color={rgb, 255:red, 0; green, 0; blue, 0 }  ,draw opacity=1 ][fill={rgb, 255:red, 74; green, 144; blue, 226 }  ,fill opacity=1 ][line width=1.5]  (244,190.5) .. controls (244,186.63) and (247.13,183.5) .. (251,183.5) .. controls (254.87,183.5) and (258,186.63) .. (258,190.5) .. controls (258,194.37) and (254.87,197.5) .. (251,197.5) .. controls (247.13,197.5) and (244,194.37) .. (244,190.5) -- cycle ; \draw  [color={rgb, 255:red, 0; green, 0; blue, 0 }  ,draw opacity=1 ][line width=1.5]  (246.05,185.55) -- (255.95,195.45) ; \draw  [color={rgb, 255:red, 0; green, 0; blue, 0 }  ,draw opacity=1 ][line width=1.5]  (255.95,185.55) -- (246.05,195.45) ;
\draw  [color={rgb, 255:red, 21; green, 2; blue, 2 }  ,draw opacity=1 ][fill={rgb, 255:red, 71; green, 142; blue, 222 }  ,fill opacity=1 ][line width=1.5]  (524,189.25) .. controls (524,185.52) and (526.91,182.5) .. (530.5,182.5) .. controls (534.09,182.5) and (537,185.52) .. (537,189.25) .. controls (537,192.98) and (534.09,196) .. (530.5,196) .. controls (526.91,196) and (524,192.98) .. (524,189.25) -- cycle ; \draw  [color={rgb, 255:red, 21; green, 2; blue, 2 }  ,draw opacity=1 ][line width=1.5]  (525.9,184.48) -- (535.1,194.02) ; \draw  [color={rgb, 255:red, 21; green, 2; blue, 2 }  ,draw opacity=1 ][line width=1.5]  (535.1,184.48) -- (525.9,194.02) ;
\draw  [fill={rgb, 255:red, 126; green, 211; blue, 33 }  ,fill opacity=1 ][line width=1.5]  (323,190.25) .. controls (323,186.52) and (326.13,183.5) .. (330,183.5) .. controls (333.87,183.5) and (337,186.52) .. (337,190.25) .. controls (337,193.98) and (333.87,197) .. (330,197) .. controls (326.13,197) and (323,193.98) .. (323,190.25) -- cycle ; \draw  [line width=1.5]  (325.05,185.48) -- (334.95,195.02) ; \draw  [line width=1.5]  (334.95,185.48) -- (325.05,195.02) ;
\draw  [fill={rgb, 255:red, 126; green, 211; blue, 33 }  ,fill opacity=1 ][line width=1.5]  (434,189.25) .. controls (434,185.52) and (437.13,182.5) .. (441,182.5) .. controls (444.87,182.5) and (448,185.52) .. (448,189.25) .. controls (448,192.98) and (444.87,196) .. (441,196) .. controls (437.13,196) and (434,192.98) .. (434,189.25) -- cycle ; \draw  [line width=1.5]  (436.05,184.48) -- (445.95,194.02) ; \draw  [line width=1.5]  (445.95,184.48) -- (436.05,194.02) ;
\draw  [color={rgb, 255:red, 126; green, 211; blue, 33 }  ,draw opacity=1 ][line width=1.5]  (326.49,226.45) -- (334.51,215.55)(324.99,216.95) -- (336.01,225.05) ;
\draw  [color={rgb, 255:red, 126; green, 211; blue, 33 }  ,draw opacity=1 ][line width=1.5]  (436.29,225.63) -- (444.71,215.37)(435.79,216.63) -- (445.21,224.37) ;
\draw  [color={rgb, 255:red, 74; green, 144; blue, 226 }  ,draw opacity=1 ][line width=1.5]  (246.21,224.93) -- (254.56,215.55)(245.83,216.19) -- (254.94,224.3) ;
\draw  [color={rgb, 255:red, 74; green, 144; blue, 226 }  ,draw opacity=1 ][line width=1.5]  (526.45,225.48) -- (534.33,216.16)(526.06,217.16) -- (534.72,224.48) ;
\draw [color={rgb, 255:red, 189; green, 16; blue, 224 }  ,draw opacity=1 ][line width=1.5]    (128,235.5) -- (580,235.5) ;
\draw  [color={rgb, 255:red, 189; green, 16; blue, 224 }  ,draw opacity=1 ][line width=1.5]  (165.17,230.15) -- (174.41,235.18) -- (166,241.5) ;
\draw  [draw opacity=0][line width=1.5]  (156.27,302.98) .. controls (152.01,297.01) and (148.1,290.67) .. (144.58,283.99) .. controls (136.33,268.33) and (130.86,251.97) .. (128,235.5) -- (280.71,212.24) -- cycle ; \draw  [color={rgb, 255:red, 189; green, 16; blue, 224 }  ,draw opacity=1 ][line width=1.5]  (156.27,302.98) .. controls (152.01,297.01) and (148.1,290.67) .. (144.58,283.99) .. controls (136.33,268.33) and (130.86,251.97) .. (128,235.5) ;  
\draw  [draw opacity=0][line width=1.5]  (580,235.5) .. controls (579.9,245.51) and (578.55,255.79) .. (575.82,266.1) .. controls (571.2,283.53) and (563.18,298.97) .. (552.9,311.59) -- (490.76,243.56) -- cycle ; \draw  [color={rgb, 255:red, 189; green, 16; blue, 224 }  ,draw opacity=1 ][line width=1.5]  (580,235.5) .. controls (579.9,245.51) and (578.55,255.79) .. (575.82,266.1) .. controls (571.2,283.53) and (563.18,298.97) .. (552.9,311.59) ;  
\draw  [color={rgb, 255:red, 189; green, 16; blue, 224 }  ,draw opacity=1 ][line width=1.5]  (135,277.5) -- (135.29,265.45) -- (145.29,272.18) ;
\draw  [color={rgb, 255:red, 189; green, 16; blue, 224 }  ,draw opacity=1 ][line width=1.5]  (578.12,280.88) -- (568.74,287.62) -- (565.17,276.64) ;

\draw (272.5,164) node   [align=left] {\begin{minipage}[lt]{19.72pt}\setlength\topsep{0pt}
\scalebox{1}{$\boldsymbol{i\delta}$}
\end{minipage}};
\draw (339,154) node [anchor=north west][inner sep=0.75pt]   [align=left] {\scalebox{1}{$\boldsymbol{(a_{+}+i\delta)}$}};
\draw (449,156) node [anchor=north west][inner sep=0.75pt]   [align=left] { \scalebox{1}{$\boldsymbol{(a_{-}+i\delta)}$}};
\draw (543,156) node [anchor=north west][inner sep=0.75pt]   [align=left] {\scalebox{1}{$\boldsymbol{(1+i\delta)}$}};
\draw (128-15,130) node [anchor=north west][inner sep=0.75pt]   [align=left] { \scalebox{1}{$\boldsymbol{\lfloor \xi}$}};

\end{tikzpicture}
    \caption{ Branch points of the $\xi$-integrand for $-\infty<\zeta<0$ and our choice of the $\xi$-contour}
    \label{first_KLT}
\end{figure}
In this regime of $\zeta$, all the branch points of the $\xi$-integrand are located in the upper half $\xi$-plane (see Fig.~\ref{first_KLT}). Consequently, we can close the $\xi$-contour in the lower half-plane, yielding $I_\xi = 0$. It then follows that the full integral $\mathcal{I}$ vanishes.

\subsection{$0<\zeta<a_{+}$} 
\begin{figure}[!htb]
    \centering

\begin{tikzpicture}[x=0.75pt,y=0.75pt,yscale=-1,xscale=1, scale=0.9]

\draw [line width=1.5]    (121,220.5) -- (598,221.49) ;
\draw [shift={(601,221.5)}, rotate = 180.12] [color={rgb, 255:red, 0; green, 0; blue, 0 }  ][line width=1.5]    (14.21,-4.28) .. controls (9.04,-1.82) and (4.3,-0.39) .. (0,0) .. controls (4.3,0.39) and (9.04,1.82) .. (14.21,4.28)   ;
\draw [line width=1.5]    (251,303.5) -- (250.02,129.5) ;
\draw [shift={(250,126.5)}, rotate = 89.68] [color={rgb, 255:red, 0; green, 0; blue, 0 }  ][line width=1.5]    (14.21,-4.28) .. controls (9.04,-1.82) and (4.3,-0.39) .. (0,0) .. controls (4.3,0.39) and (9.04,1.82) .. (14.21,4.28)   ;
\draw  [color={rgb, 255:red, 0; green, 0; blue, 0 }  ,draw opacity=1 ][fill={rgb, 255:red, 74; green, 144; blue, 226 }  ,fill opacity=1 ][line width=1.5]  (244,249.5) .. controls (244,245.63) and (247.13,242.5) .. (251,242.5) .. controls (254.87,242.5) and (258,245.63) .. (258,249.5) .. controls (258,253.37) and (254.87,256.5) .. (251,256.5) .. controls (247.13,256.5) and (244,253.37) .. (244,249.5) -- cycle ; \draw  [color={rgb, 255:red, 0; green, 0; blue, 0 }  ,draw opacity=1 ][line width=1.5]  (246.05,244.55) -- (255.95,254.45) ; \draw  [color={rgb, 255:red, 0; green, 0; blue, 0 }  ,draw opacity=1 ][line width=1.5]  (255.95,244.55) -- (246.05,254.45) ;
\draw  [color={rgb, 255:red, 21; green, 2; blue, 2 }  ,draw opacity=1 ][fill={rgb, 255:red, 71; green, 142; blue, 222 }  ,fill opacity=1 ][line width=1.5]  (524,189.25) .. controls (524,185.52) and (526.91,182.5) .. (530.5,182.5) .. controls (534.09,182.5) and (537,185.52) .. (537,189.25) .. controls (537,192.98) and (534.09,196) .. (530.5,196) .. controls (526.91,196) and (524,192.98) .. (524,189.25) -- cycle ; \draw  [color={rgb, 255:red, 21; green, 2; blue, 2 }  ,draw opacity=1 ][line width=1.5]  (525.9,184.48) -- (535.1,194.02) ; \draw  [color={rgb, 255:red, 21; green, 2; blue, 2 }  ,draw opacity=1 ][line width=1.5]  (535.1,184.48) -- (525.9,194.02) ;
\draw  [fill={rgb, 255:red, 126; green, 211; blue, 33 }  ,fill opacity=1 ][line width=1.5]  (323,190.25) .. controls (323,186.52) and (326.13,183.5) .. (330,183.5) .. controls (333.87,183.5) and (337,186.52) .. (337,190.25) .. controls (337,193.98) and (333.87,197) .. (330,197) .. controls (326.13,197) and (323,193.98) .. (323,190.25) -- cycle ; \draw  [line width=1.5]  (325.05,185.48) -- (334.95,195.02) ; \draw  [line width=1.5]  (334.95,185.48) -- (325.05,195.02) ;
\draw  [fill={rgb, 255:red, 126; green, 211; blue, 33 }  ,fill opacity=1 ][line width=1.5]  (434,189.25) .. controls (434,185.52) and (437.13,182.5) .. (441,182.5) .. controls (444.87,182.5) and (448,185.52) .. (448,189.25) .. controls (448,192.98) and (444.87,196) .. (441,196) .. controls (437.13,196) and (434,192.98) .. (434,189.25) -- cycle ; \draw  [line width=1.5]  (436.05,184.48) -- (445.95,194.02) ; \draw  [line width=1.5]  (445.95,184.48) -- (436.05,194.02) ;
\draw  [color={rgb, 255:red, 126; green, 211; blue, 33 }  ,draw opacity=1 ][line width=1.5]  (326.49,226.45) -- (334.51,215.55)(324.99,216.95) -- (336.01,225.05) ;
\draw  [color={rgb, 255:red, 126; green, 211; blue, 33 }  ,draw opacity=1 ][line width=1.5]  (436.29,225.63) -- (444.71,215.37)(435.79,216.63) -- (445.21,224.37) ;
\draw  [color={rgb, 255:red, 74; green, 144; blue, 226 }  ,draw opacity=1 ][line width=1.5]  (246.21,224.93) -- (254.56,215.55)(245.83,216.19) -- (254.94,224.3) ;
\draw  [color={rgb, 255:red, 74; green, 144; blue, 226 }  ,draw opacity=1 ][line width=1.5]  (526.45,225.48) -- (534.33,216.16)(526.06,217.16) -- (534.72,224.48) ;
\draw [color={rgb, 255:red, 189; green, 16; blue, 224 }  ,draw opacity=1 ][line width=1.5]    (120,235.5) -- (261,236.5) ;
\draw [color={rgb, 255:red, 189; green, 16; blue, 224 }  ,draw opacity=1 ][line width=1.5]    (120,264.5) -- (261,265.5) ;
\draw [color={rgb, 255:red, 189; green, 16; blue, 224 }  ,draw opacity=1 ][line width=1.5]    (261,236.5) .. controls (280,236.5) and (283,265.5) .. (261,265.5) ;
\draw  [color={rgb, 255:red, 189; green, 16; blue, 224 }  ,draw opacity=1 ][line width=1.5]  (165.17,230.15) -- (174.41,235.18) -- (166,241.5) ;
\draw  [color={rgb, 255:red, 189; green, 16; blue, 224 }  ,draw opacity=1 ][line width=1.5]  (174,270.5) -- (167,264.25) -- (174,258) ;
\draw  [color={rgb, 255:red, 189; green, 16; blue, 224 }  ,draw opacity=1 ][line width=1.5]  (280.4,243.85) -- (276.73,252.12) -- (270.14,245.92) ;

\draw (266.5,274.25+3) node   [align=left] {\begin{minipage}[lt]{17pt}\setlength\topsep{0pt}
\scalebox{1}{$\boldsymbol{i\delta}$}
\end{minipage}};
\draw (320+15,155) node [anchor=north west][inner sep=0.75pt]   [align=left] {\scalebox{1}{$\boldsymbol{(a_{+}+i\delta)}$}};
\draw (449,156) node [anchor=north west][inner sep=0.75pt]   [align=left] {\scalebox{1}{$\boldsymbol{(a_{-}+i\delta)}$}};
\draw (543,156) node [anchor=north west][inner sep=0.75pt]   [align=left] {\scalebox{1}{$\boldsymbol{(1+i\delta)}$}};
\draw (90,226) node [anchor=north west][inner sep=0.75pt]   [align=left] { \scalebox{1}{$\boldsymbol{e^{i\pi}}$} };
\draw (89-5,256) node [anchor=north west][inner sep=0.75pt]   [align=left] { \scalebox{1}{$\boldsymbol{e^{-i\pi}}$} };
\draw (128-30,130) node [anchor=north west][inner sep=0.75pt]   [align=left] { \scalebox{1}{$\boldsymbol{\lfloor \xi}$}};

\end{tikzpicture}
    
    \caption{ Branch points of the $\xi$-integrand for $0<\zeta<a_{+}$ and our choice of the $\xi$-contour}
    \label{second_KLT}
\end{figure}
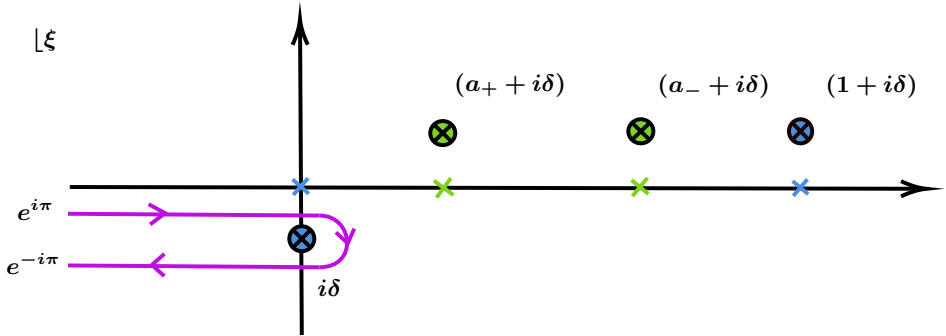

The $\zeta$-integral now takes the form
\begin{equation}
    \mathcal{I}_\zeta^{(0,a_{+})} = \int_{0}^{a_{+}} d\zeta \, \zeta^{-s/4-b-1} (1-\zeta)^{-t/4-b-1} (a_{+}-\zeta)^{q/2} (a_{-}-\zeta )^{q/2} \, ,
\end{equation}
Note that we have chosen $(a_{+}-\zeta)^{q/2} (a_{-}-\zeta )^{q/2}$ instead of $(\zeta-a_{+})^{q/2} (\zeta-a_{-} )^{q/2}$ in the $\zeta$-integrand, we have to make the corresponding choice in the $\xi$-integrand. The branch point $\xi_1$ now moves into the lower half plane, and we can close the $\xi$-contour around it as illustrated in Fig.~\ref{second_KLT}. 

We introduce the shifted variable $\xi'=\xi-i\delta$. Along the segment $\int_{-\infty}^{0}d\xi'$, we have $\xi'=(-\xi')\,e^{i\pi}$, while along $\int_0^{-\infty}d\xi'$, we have $\xi'=(-\xi')\, e^{-i\pi}$. This gives
\begin{equation}
    \mathcal{I}_\xi^{(0,a_{+})} = 2i\sin[\pi(-s/4-b-1)]\int_{-\infty}^{0}d\xi' \, (-\xi')^{-s/4-b-1} (1-\xi')^{-t/4-b-1} (a_{+}-\xi')^{q/2} (a_{-}-\xi')^{q/2} \, .
\end{equation}
Making suitable change of variables, we can rephrase the integrals in terms of Appell's $\mathds{F}_1$ function:
\begin{equation}\label{first_domain}
    \begin{split}
        \mathcal{I}_\zeta^{(0,a_{+})} &= a_{+}^{ -\frac{s}{4}-b+\frac{q}{2}} a_{-}^{\frac{q}{2}} \,\mathds{B}\Big(-\frac{s}{4}-b, 1+\frac{q}{2}\Big) \, \mathds{F}_1\Big( -\frac{s}{4}-b, \frac{t}{4}+b+1, -\frac{q}{2},1-b+\frac{q}{2} -\frac{s}{4} \Big| a_{+}, \frac{a_{+}}{a_{-}} \Big) \, ,\\
        \mathcal{I}_\xi^{(0,a_{+})} &= 2i\sin\left[\pi(s/4+b) \right] \mathds{B} \left( \frac{s+t}{4}+2b+1-q, -\frac{s}{4}-b \right) \\
        &  \times \mathds{F}_1 \left( \frac{s+t}{4}+2b+1-q,-\frac{q}{2},-\frac{q}{2}, \frac{t}{4}+b+1-q \Big| 1-a_{+}, 1-a_{-} \right) \\
        &=2i \sin [\pi (s/4+b)]\mathds{B}\left(-\frac{s}{4}-b+\frac{q}{2},\frac{s+t}{4}+2 b-q+1\right)\, {}_3F_2\left[\substack{\frac{s+t}{8}+b+\frac{1-q}{2},\frac{s+t}{8}+b+1-\frac{q}{2},-\frac{q}{2}\\ \frac{t}{4}+b+1-\frac{q}{2},\frac{s}{4}+b+1-\frac{q}{2} };4 a_{+}a_{-}\right]\,,
    \end{split}
\end{equation}
where we used the identity
\begin{align}
\mathds{F}_{1}(a,b,b,c\Big|x,1-x)
=\frac{\Gamma(c)\Gamma(c-b-a)}{\Gamma(c-a)\Gamma(c-b)}{}_3F_{2}\left[\substack{\frac{a}{2},\frac{a+1}{2},b\\ c-b,1+a+b-c};4x(1-x)\right]\,.
\end{align}

\subsection{ $a_{+}<\zeta<a_{-}$} 
\begin{figure}[!htb]
    \centering

\begin{tikzpicture}[x=0.75pt,y=0.75pt,yscale=-1,xscale=1, scale=0.9]

\draw [line width=1.5]    (121,220.5) -- (598,221.49) ;
\draw [shift={(601,221.5)}, rotate = 180.12] [color={rgb, 255:red, 0; green, 0; blue, 0 }  ][line width=1.5]    (14.21,-4.28) .. controls (9.04,-1.82) and (4.3,-0.39) .. (0,0) .. controls (4.3,0.39) and (9.04,1.82) .. (14.21,4.28)   ;
\draw [line width=1.5]    (251,303.5) -- (250.02,129.5) ;
\draw [shift={(250,126.5)}, rotate = 89.68] [color={rgb, 255:red, 0; green, 0; blue, 0 }  ][line width=1.5]    (14.21,-4.28) .. controls (9.04,-1.82) and (4.3,-0.39) .. (0,0) .. controls (4.3,0.39) and (9.04,1.82) .. (14.21,4.28)   ;
\draw  [color={rgb, 255:red, 0; green, 0; blue, 0 }  ,draw opacity=1 ][fill={rgb, 255:red, 74; green, 144; blue, 226 }  ,fill opacity=1 ][line width=1.5]  (244,249.5) .. controls (244,245.63) and (247.13,242.5) .. (251,242.5) .. controls (254.87,242.5) and (258,245.63) .. (258,249.5) .. controls (258,253.37) and (254.87,256.5) .. (251,256.5) .. controls (247.13,256.5) and (244,253.37) .. (244,249.5) -- cycle ; \draw  [color={rgb, 255:red, 0; green, 0; blue, 0 }  ,draw opacity=1 ][line width=1.5]  (246.05,244.55) -- (255.95,254.45) ; \draw  [color={rgb, 255:red, 0; green, 0; blue, 0 }  ,draw opacity=1 ][line width=1.5]  (255.95,244.55) -- (246.05,254.45) ;
\draw  [color={rgb, 255:red, 21; green, 2; blue, 2 }  ,draw opacity=1 ][fill={rgb, 255:red, 71; green, 142; blue, 222 }  ,fill opacity=1 ][line width=1.5]  (524,189.25) .. controls (524,185.52) and (526.91,182.5) .. (530.5,182.5) .. controls (534.09,182.5) and (537,185.52) .. (537,189.25) .. controls (537,192.98) and (534.09,196) .. (530.5,196) .. controls (526.91,196) and (524,192.98) .. (524,189.25) -- cycle ; \draw  [color={rgb, 255:red, 21; green, 2; blue, 2 }  ,draw opacity=1 ][line width=1.5]  (525.9,184.48) -- (535.1,194.02) ; \draw  [color={rgb, 255:red, 21; green, 2; blue, 2 }  ,draw opacity=1 ][line width=1.5]  (535.1,184.48) -- (525.9,194.02) ;
\draw  [fill={rgb, 255:red, 126; green, 211; blue, 33 }  ,fill opacity=1 ][line width=1.5]  (323,250.25) .. controls (323,246.52) and (326.13,243.5) .. (330,243.5) .. controls (333.87,243.5) and (337,246.52) .. (337,250.25) .. controls (337,253.98) and (333.87,257) .. (330,257) .. controls (326.13,257) and (323,253.98) .. (323,250.25) -- cycle ; \draw  [line width=1.5]  (325.05,245.48) -- (334.95,255.02) ; \draw  [line width=1.5]  (334.95,245.48) -- (325.05,255.02) ;
\draw  [fill={rgb, 255:red, 126; green, 211; blue, 33 }  ,fill opacity=1 ][line width=1.5]  (434,189.25) .. controls (434,185.52) and (437.13,182.5) .. (441,182.5) .. controls (444.87,182.5) and (448,185.52) .. (448,189.25) .. controls (448,192.98) and (444.87,196) .. (441,196) .. controls (437.13,196) and (434,192.98) .. (434,189.25) -- cycle ; \draw  [line width=1.5]  (436.05,184.48) -- (445.95,194.02) ; \draw  [line width=1.5]  (445.95,184.48) -- (436.05,194.02) ;
\draw  [color={rgb, 255:red, 126; green, 211; blue, 33 }  ,draw opacity=1 ][line width=1.5]  (326.49,226.45) -- (334.51,215.55)(324.99,216.95) -- (336.01,225.05) ;
\draw  [color={rgb, 255:red, 126; green, 211; blue, 33 }  ,draw opacity=1 ][line width=1.5]  (436.29,225.63) -- (444.71,215.37)(435.79,216.63) -- (445.21,224.37) ;
\draw  [color={rgb, 255:red, 74; green, 144; blue, 226 }  ,draw opacity=1 ][line width=1.5]  (246.21,224.93) -- (254.56,215.55)(245.83,216.19) -- (254.94,224.3) ;
\draw  [color={rgb, 255:red, 74; green, 144; blue, 226 }  ,draw opacity=1 ][line width=1.5]  (526.45,225.48) -- (534.33,216.16)(526.06,217.16) -- (534.72,224.48) ;
\draw [color={rgb, 255:red, 189; green, 16; blue, 224 }  ,draw opacity=1 ][line width=1.5]    (121,235.5) -- (334,235.5) ;
\draw [color={rgb, 255:red, 189; green, 16; blue, 224 }  ,draw opacity=1 ][line width=1.5]    (123,264.5) -- (334,264.5) ;
\draw [color={rgb, 255:red, 189; green, 16; blue, 224 }  ,draw opacity=1 ][line width=1.5]    (334,235.5) .. controls (353,235.5) and (356,264.5) .. (334,264.5) ;
\draw  [color={rgb, 255:red, 189; green, 16; blue, 224 }  ,draw opacity=1 ][line width=1.5]  (165.17,230.15) -- (174.41,235.18) -- (166,241.5) ;
\draw  [color={rgb, 255:red, 189; green, 16; blue, 224 }  ,draw opacity=1 ][line width=1.5]  (174,270.5) -- (167,264.25) -- (174,258) ;
\draw  [color={rgb, 255:red, 189; green, 16; blue, 224 }  ,draw opacity=1 ][line width=1.5]  (353.82,244.28) -- (349.39,252.17) -- (343.41,245.38) ;

\draw (266.5,274.25+2) node   [align=left] {\begin{minipage}[lt]{17pt}\setlength\topsep{0pt}
{\scalebox{1}{$\boldsymbol{i\delta}$}}
\end{minipage}};
\draw (320,270) node [anchor=north west][inner sep=0.75pt]   [align=left] {\scalebox{1}{$\boldsymbol{(a_{+}+i\delta)}$}};
\draw (449,156) node [anchor=north west][inner sep=0.75pt]   [align=left] {\scalebox{1}{$\boldsymbol{(a_{-}+i\delta)}$}};
\draw (543,156) node [anchor=north west][inner sep=0.75pt]   [align=left] {\scalebox{1}{$\boldsymbol{(1+i\delta)}$}};
\draw (90,226) node [anchor=north west][inner sep=0.75pt]   [align=left] {\scalebox{1}{$\boldsymbol{e^{i\pi}}$}};
\draw (93-5,255) node [anchor=north west][inner sep=0.75pt]   [align=left] {\scalebox{1}{$\boldsymbol{e^{-i\pi}}$}};
\draw (128-30,130) node [anchor=north west][inner sep=0.75pt]   [align=left] { \scalebox{1}{$\boldsymbol{\lfloor \xi}$}};

\end{tikzpicture}

\caption{Branch points of the $\xi$-integrand for $a_{+}<\zeta<a_{-}$ and our choice of the $\xi$-contour}
\label{third_KLT}
\end{figure}
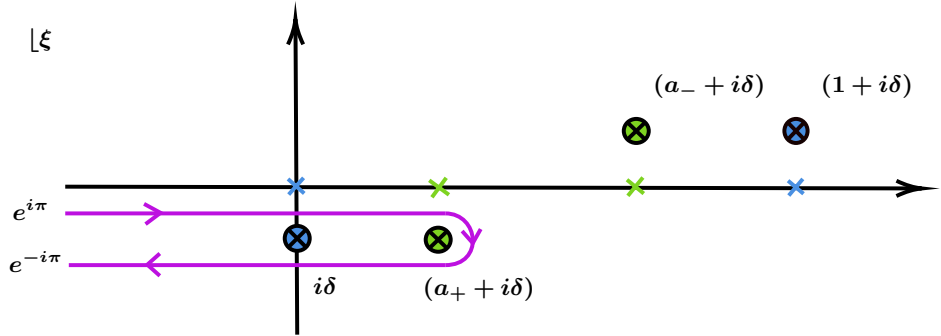

In this regime of $\zeta$, the integrals take the form
\begin{equation}
\begin{split}
    \mathcal{I}_\zeta^{(a_{+},a_{-})} &=\int_{a_{+}}^{a_{-}} d\zeta\, \zeta^{-s/4-b-1} (1-\zeta)^{-t/4-b-1} (\zeta-a_{+})^{q/2} (a_{-}-\zeta)^{q/2} \, , \\
     \mathcal{I}_\xi^{(a_{+},a_{-})}  &=  \int d\xi  \, (\xi-i\delta)^{- s/4-b-1} (1-\xi+i\delta)^{-t/4-b-1}  (\xi-a_{+}-i\delta)^{q/2} (a_{-}+i\delta-\xi)^{q/2} \, .
\end{split}    
\end{equation}
The branch points $\xi_1$ and $\xi_2$ now lie below the real $\xi$-axis, while $\xi_3$ and $\xi_4$ lie above. The $\xi$-contour can be closed in either half-plane, with both choices yielding the same result. We choose to close the contour in the lower half-plane, as shown in Fig.~\ref{third_KLT}.

Evaluating the contour then gives 
\begin{equation}
    \begin{split}
        \mathcal{I}_\xi^{(a_{+},a_{-})}=& 2i\sin[\pi q/2]\int_{0}^{a_{+}}d\xi\, \xi^{-s/4-b-1} (1-\xi)^{-t/4-b-1} (a_{+}-\xi)^{q/2} (a_{-}-\xi)^{q/2}\\
        +& 2i\sin[\pi (- s/4-b-1+q/2)] \int_{-\infty}^{0}d\xi\, (-\xi)^{-s/4-b-1} (1-\xi)^{-t/4-b-1} (a_{+}-\xi)^{q/2} (a_{-}-\xi)^{q/2} \, .
    \end{split}
\end{equation}
After an appropriate change of variables, the integrals take the form in terms of Appell's $\mathds{F}_1$ function:
\begin{equation}\label{second_domain}
    \begin{split}
    \mathcal{I}_\zeta^{(a_{+},a_{-})}&= (a_{-}-a_{+})^{1+q} \,a_{+}^{-\frac{s}{4}-b-1} a_{-}^{ -\frac{t}{4}-b-1} \mathds{B}\left(1+\frac{q}{2}, 1+\frac{q}{2} \right)
\mathds{F}_1 \left( 1+\frac{q}{2}, \frac{s}{4}+b+1, \frac{t}{4}+b+1, 2+q \Big| \frac{a_{+}-a_{-}}{a_{+}}, \frac{a_{-} -a_{+}}{a_{-}} \right) \,\\
        \mathcal{I}_\xi^{(a_{+},a_{-})} &= 2i\sin\left( \frac{\pi q}{2}\right) \mathds{B}\left( -\frac{s}{4}-b,1+\frac{q}{2}\right)a_+^{-\frac{s}{4}-b+\frac{q}{2}} a_{-}^{\frac{q}{2}} \mathds{F}_1\left( -\frac{s}{4}-b, \frac{t}{4}+b+1,-\frac{q}{2}, 1-b+\frac{q}{2}-\frac{s}{4} \Big| a_{+}, \frac{a_{+}}{a_{-}}  \right)\\
         & \hspace{-50pt}+ 2i \sin\left[ \pi\left(\frac{s}{4}+b-\frac{q}{2} \right) \right] \mathds{B} \left( \frac{s+t}{4}+2b+1-q, -\frac{s}{4}-b \right)  \mathds{F}_1\left(\frac{s+t}{4}+2b+1-q, -\frac{q}{2}, -\frac{q}{2}, \frac{t}{4}+b+1-q \Big| a_+,a_-  \right) \, ,\\
        &=2i\sin\left( \frac{\pi q}{2}\right) a_+^{-\frac{s}{4}-b+\frac{q}{2}} a_{-}^{\frac{q}{2}} \mathds{B}\left( -\frac{s}{4}-b,1+\frac{q}{2}\right)\mathds{F}_1\left( -\frac{s}{4}-b, \frac{t}{4}+b+1,-\frac{q}{2}, 1-b+\frac{q}{2}-\frac{s}{4} \Big| a_{+}, \frac{a_{+}}{a_{-}}  \right)\\
         &+2i \sin\left[ \pi\left(\frac{s}{4}+b-\frac{q}{2} \right) \right]\mathds{B}\left(-\frac{s}{4}-b+\frac{q}{2},\frac{s+t}{4}+2 b-q+1\right)\, {}_3F_2\left[\substack{\frac{s+t}{8}+b+\frac{1-q}{2},\frac{s+t}{8}+b+1-\frac{q}{2},-\frac{q}{2}\\ \frac{t}{4}+b+1-\frac{q}{2},\frac{s}{4}+b+1-\frac{q}{2}};4 a_{+}a_{-}\right]\, .
    \end{split}
\end{equation}

\subsection{$a_{-}<\zeta<1$} 
In this regime, the $\zeta$-integral becomes
\begin{equation}
    \mathcal{I}_\zeta^{(a_{-},1)} = \int_{a_{-}}^1 d\zeta \, \zeta^{-s/4-b-1} (1-\zeta)^{-t/4-b-1} (\zeta-a_{+})^{q/2} (\zeta-a_{-})^{q/2}  \, ,
\end{equation}
In this case only the branch point $\xi_4$ lies above the real $\xi$-axis, while all others are located below. The $\xi$-contour can therefore be closed around $\xi_4$, as shown in Fig.~\ref{fourth_KLT}.

\begin{figure}[!htb]
    \centering
 \begin{tikzpicture}[x=0.75pt,y=0.75pt,yscale=-1,xscale=1, scale=0.9]

\draw [line width=1.5]    (121,220.5) -- (598,221.49) ;
\draw [shift={(601,221.5)}, rotate = 180.12] [color={rgb, 255:red, 0; green, 0; blue, 0 }  ][line width=1.5]    (14.21,-4.28) .. controls (9.04,-1.82) and (4.3,-0.39) .. (0,0) .. controls (4.3,0.39) and (9.04,1.82) .. (14.21,4.28)   ;
\draw [line width=1.5]    (251,303.5) -- (250.02,129.5) ;
\draw [shift={(250,126.5)}, rotate = 89.68] [color={rgb, 255:red, 0; green, 0; blue, 0 }  ][line width=1.5]    (14.21,-4.28) .. controls (9.04,-1.82) and (4.3,-0.39) .. (0,0) .. controls (4.3,0.39) and (9.04,1.82) .. (14.21,4.28)   ;
\draw  [color={rgb, 255:red, 0; green, 0; blue, 0 }  ,draw opacity=1 ][fill={rgb, 255:red, 74; green, 144; blue, 226 }  ,fill opacity=1 ][line width=1.5]  (244,249.5) .. controls (244,245.63) and (247.13,242.5) .. (251,242.5) .. controls (254.87,242.5) and (258,245.63) .. (258,249.5) .. controls (258,253.37) and (254.87,256.5) .. (251,256.5) .. controls (247.13,256.5) and (244,253.37) .. (244,249.5) -- cycle ; \draw  [color={rgb, 255:red, 0; green, 0; blue, 0 }  ,draw opacity=1 ][line width=1.5]  (246.05,244.55) -- (255.95,254.45) ; \draw  [color={rgb, 255:red, 0; green, 0; blue, 0 }  ,draw opacity=1 ][line width=1.5]  (255.95,244.55) -- (246.05,254.45) ;
\draw  [color={rgb, 255:red, 21; green, 2; blue, 2 }  ,draw opacity=1 ][fill={rgb, 255:red, 71; green, 142; blue, 222 }  ,fill opacity=1 ][line width=1.5]  (524,189.25) .. controls (524,185.52) and (526.91,182.5) .. (530.5,182.5) .. controls (534.09,182.5) and (537,185.52) .. (537,189.25) .. controls (537,192.98) and (534.09,196) .. (530.5,196) .. controls (526.91,196) and (524,192.98) .. (524,189.25) -- cycle ; \draw  [color={rgb, 255:red, 21; green, 2; blue, 2 }  ,draw opacity=1 ][line width=1.5]  (525.9,184.48) -- (535.1,194.02) ; \draw  [color={rgb, 255:red, 21; green, 2; blue, 2 }  ,draw opacity=1 ][line width=1.5]  (535.1,184.48) -- (525.9,194.02) ;
\draw  [fill={rgb, 255:red, 126; green, 211; blue, 33 }  ,fill opacity=1 ][line width=1.5]  (323,250.25) .. controls (323,246.52) and (326.13,243.5) .. (330,243.5) .. controls (333.87,243.5) and (337,246.52) .. (337,250.25) .. controls (337,253.98) and (333.87,257) .. (330,257) .. controls (326.13,257) and (323,253.98) .. (323,250.25) -- cycle ; \draw  [line width=1.5]  (325.05,245.48) -- (334.95,255.02) ; \draw  [line width=1.5]  (334.95,245.48) -- (325.05,255.02) ;
\draw  [fill={rgb, 255:red, 126; green, 211; blue, 33 }  ,fill opacity=1 ][line width=1.5]  (434,250.25) .. controls (434,246.52) and (437.13,243.5) .. (441,243.5) .. controls (444.87,243.5) and (448,246.52) .. (448,250.25) .. controls (448,253.98) and (444.87,257) .. (441,257) .. controls (437.13,257) and (434,253.98) .. (434,250.25) -- cycle ; \draw  [line width=1.5]  (436.05,245.48) -- (445.95,255.02) ; \draw  [line width=1.5]  (445.95,245.48) -- (436.05,255.02) ;
\draw  [color={rgb, 255:red, 126; green, 211; blue, 33 }  ,draw opacity=1 ][line width=1.5]  (326.49,226.45) -- (334.51,215.55)(324.99,216.95) -- (336.01,225.05) ;
\draw  [color={rgb, 255:red, 126; green, 211; blue, 33 }  ,draw opacity=1 ][line width=1.5]  (436.29,225.63) -- (444.71,215.37)(435.79,216.63) -- (445.21,224.37) ;
\draw  [color={rgb, 255:red, 74; green, 144; blue, 226 }  ,draw opacity=1 ][line width=1.5]  (246.21,224.93) -- (254.56,215.55)(245.83,216.19) -- (254.94,224.3) ;
\draw  [color={rgb, 255:red, 74; green, 144; blue, 226 }  ,draw opacity=1 ][line width=1.5]  (526.45,225.48) -- (534.33,216.16)(526.06,217.16) -- (534.72,224.48) ;
\draw [color={rgb, 255:red, 189; green, 16; blue, 224 }  ,draw opacity=1 ][line width=1.5]    (520,174.5) -- (595,175.5) ;
\draw [color={rgb, 255:red, 189; green, 16; blue, 224 }  ,draw opacity=1 ][line width=1.5]    (520,205.5) -- (596,205.5) ;
\draw [color={rgb, 255:red, 189; green, 16; blue, 224 }  ,draw opacity=1 ][line width=1.5]    (520,205.5) .. controls (502,205.5) and (503,175.5) .. (520,174.5) ;
\draw  [color={rgb, 255:red, 189; green, 16; blue, 224 }  ,draw opacity=1 ][line width=1.5]  (565.82,199.78) -- (574.19,204.98) -- (566.19,210.73) ;
\draw  [color={rgb, 255:red, 189; green, 16; blue, 224 }  ,draw opacity=1 ][line width=1.5]  (577.15,180.28) -- (568.85,175.05) -- (576.86,169.38) ;
\draw  [color={rgb, 255:red, 189; green, 16; blue, 224 }  ,draw opacity=1 ][line width=1.5]  (513.41,185.41) -- (506.4,191.14) -- (502.5,182.97) ;

\draw (266.5,274.25) node   [align=left] {\begin{minipage}[lt]{17pt}\setlength\topsep{0pt}
{\scalebox{1}{$\boldsymbol{i\delta}$}}
\end{minipage}};
\draw (320,270) node [anchor=north west][inner sep=0.75pt]   [align=left] {\scalebox{1}{$\boldsymbol{(a_{+}+i\delta)}$}};
\draw (429,269) node [anchor=north west][inner sep=0.75pt]   [align=left] {\scalebox{1}{$\boldsymbol{(a_{-}+i\delta)}$}};
\draw (539-20,100+50) node [anchor=north west][inner sep=0.75pt]   [align=left] {\scalebox{1}{$\boldsymbol{(1+i\delta)}$}};
\draw (601,168-2) node [anchor=north west][inner sep=0.75pt]   [align=left] {\scalebox{1}{$\boldsymbol{e^{-i\pi}}$}};
\draw (601+2,196) node [anchor=north west][inner sep=0.75pt]   [align=left] {\scalebox{1}{$\boldsymbol{e^{i\pi}}$}};
\draw (128-30,130) node [anchor=north west][inner sep=0.75pt]   [align=left] { \scalebox{1}{$\boldsymbol{\lfloor \xi}$}};

\end{tikzpicture}
    \caption{Branch points of the $\xi$-integrand for $a_{-}<\zeta<1$ and our choice of the $\xi$-contour}
    \label{fourth_KLT}
\end{figure}
Introducing the shifted variable $\xi'=\xi-(1+i\delta)$, we note that along $\int_0^\infty d\xi'$, one has $1-\xi+i\delta=-\xi'=e^{i\pi}\xi'$, while along $\int_{\infty}^0 d\xi'$, it becomes $1-\xi+i\delta=-\xi'=e^{-i\pi}\xi'$. With this prescription, the $\xi$-contour integral evaluates to
\begin{equation}
    \mathcal{I}_\xi^{(a_{-},1)} = -2i\sin[\pi(t/4+b+1)] \int_0^\infty d\xi' \, (1+\xi')^{-s/4-b-1} (\xi')^{-t/4-b-1} (\xi'+1-a_{+})^{q/2} (\xi'+1-a_{-})^{q/2} \, .
\end{equation}
By performing a suitable change of variables, the integrals can be expressed in terms of Appell’s $\mathds{F}_1$:
\begin{equation}\label{third_domain}
    \begin{split}
        \mathcal{I}_\zeta^{(a_{-},1)} &= a_{+}^{-\frac{t}{4}-b+\frac{q}{2} } a_{-}^{ -\frac{s}{4}-b-1}(a_{-}-a_{+})^{\frac{q}{2}} \mathds{B}\left(1+\frac{q}{2}, -\frac{t}{4}-b \right)\\
        & \times \mathds{F}_1 \left( 1+\frac{q}{2}, \frac{s}{4}+b+1,-\frac{q}{2},1+\frac{q}{2}-b-\frac{t}{4} \Big| \frac{a_{-}-1}{a_{-}}, \frac{a_{-}-1}{a_{-}-a_{+}} \right)  \, ,\\
        &=a_{+}^{-\frac{t}{4}-b+\frac{q}{2} }a_{-}^{\frac{q}{2}}\mathds{B}\left(1+\frac{q}{2}, -\frac{t}{4}-b \right)
\mathds{F}_1 \left(\frac{t}{4}+b, \frac{s}{4}+b+1,-\frac{q}{2},1+\frac{q}{2}-b-\frac{t}{4} \Big| a_{+}, \frac{a_{+}}{a_{-}} \right) 
        \\ 
        \mathcal{I}_\xi^{(a_{-},1)} &= 2i \sin\left[\pi(t/4+b ) \right] \mathds{B}\left( \frac{s+t}{4}+2b-q+1, -\frac{t}{4}-b \right) \\
        & \quad\quad\times \mathds{F}_1\left(\frac{s+t}{4}+2b-q+1, -\frac{q}{2}, -\frac{q}{2}, \frac{s}{4}+b+1-q \Big| a_{+},a_{-}  \right) \, \\
        &=2i \sin [\pi (t/4+b)]\mathds{B}\left(-\frac{t}{4}-b+\frac{q}{2},\frac{s+t}{4}+2 b-q+1\right)\, {}_3F_2\left[\substack{\frac{s+t}{8}+b+\frac{1-q}{2},\frac{s+t}{8}+b+1-\frac{q}{2},-\frac{q}{2}\\ \frac{s}{4}+b+1-\frac{q}{2},\frac{t}{4}+b+1-\frac{q}{2}};4 a_{+}a_{-}\right]
    \end{split}
\end{equation}

\subsection{$1<\zeta<\infty$} In this regime of $\zeta$, all branch points of the $\xi$-integrand lie below the real $\xi$-axis (see Fig.~\ref{fifth_KLT}), and we can close the $\xi$-contour in the upper half plane, yielding $\mathcal{I}_\xi=0$.  Consequently, the full integral $\mathcal{I}$ vanishes.

\subsection{Full answer}
The resulting expression for the four-point closed-string amplitude \eqref{closed_string_amplitude} takes the form 
\begin{equation}
    \mathcal{A}^{ \text{closed}}_4/\tilde{\mathcal{N}}_4^{\rm closed}= i \left[ \mathcal{I}_\zeta^{(0,a_{+})}\, \mathcal{I}_\xi^{(0,a_{+})} + \mathcal{I}_\zeta^{(a_{+},a_{-})}\, \mathcal{I}_\xi^{(a_{+},a_{-})}+\mathcal{I}_\zeta^{(a_{-},1)}\, \mathcal{I}_\xi^{(a_{-},1)} \right]\, 
\end{equation}
where the explicit forms of $\mathcal{I}_\zeta$ and $\mathcal{I}_\xi$ in each integration domain are given in Eqs.~\eqref{first_domain}, \eqref{second_domain}, and \eqref{third_domain}.

\begin{figure}
    \centering
\begin{tikzpicture}[x=0.75pt,y=0.75pt,yscale=-1,xscale=1, scale=0.9]

\draw [line width=1.5]    (121,220.5) -- (598,221.49) ;
\draw [shift={(601,221.5)}, rotate = 180.12] [color={rgb, 255:red, 0; green, 0; blue, 0 }  ][line width=1.5]    (14.21,-4.28) .. controls (9.04,-1.82) and (4.3,-0.39) .. (0,0) .. controls (4.3,0.39) and (9.04,1.82) .. (14.21,4.28)   ;
\draw [line width=1.5]    (251,303.5) -- (250.02,129.5) ;
\draw [shift={(250,126.5)}, rotate = 89.68] [color={rgb, 255:red, 0; green, 0; blue, 0 }  ][line width=1.5]    (14.21,-4.28) .. controls (9.04,-1.82) and (4.3,-0.39) .. (0,0) .. controls (4.3,0.39) and (9.04,1.82) .. (14.21,4.28)   ;
\draw  [color={rgb, 255:red, 0; green, 0; blue, 0 }  ,draw opacity=1 ][fill={rgb, 255:red, 74; green, 144; blue, 226 }  ,fill opacity=1 ][line width=1.5]  (244,249.5) .. controls (244,245.63) and (247.13,242.5) .. (251,242.5) .. controls (254.87,242.5) and (258,245.63) .. (258,249.5) .. controls (258,253.37) and (254.87,256.5) .. (251,256.5) .. controls (247.13,256.5) and (244,253.37) .. (244,249.5) -- cycle ; \draw  [color={rgb, 255:red, 0; green, 0; blue, 0 }  ,draw opacity=1 ][line width=1.5]  (246.05,244.55) -- (255.95,254.45) ; \draw  [color={rgb, 255:red, 0; green, 0; blue, 0 }  ,draw opacity=1 ][line width=1.5]  (255.95,244.55) -- (246.05,254.45) ;
\draw  [color={rgb, 255:red, 21; green, 2; blue, 2 }  ,draw opacity=1 ][fill={rgb, 255:red, 71; green, 142; blue, 222 }  ,fill opacity=1 ][line width=1.5]  (523,250.25) .. controls (523,246.52) and (525.91,243.5) .. (529.5,243.5) .. controls (533.09,243.5) and (536,246.52) .. (536,250.25) .. controls (536,253.98) and (533.09,257) .. (529.5,257) .. controls (525.91,257) and (523,253.98) .. (523,250.25) -- cycle ; \draw  [color={rgb, 255:red, 21; green, 2; blue, 2 }  ,draw opacity=1 ][line width=1.5]  (524.9,245.48) -- (534.1,255.02) ; \draw  [color={rgb, 255:red, 21; green, 2; blue, 2 }  ,draw opacity=1 ][line width=1.5]  (534.1,245.48) -- (524.9,255.02) ;
\draw  [fill={rgb, 255:red, 126; green, 211; blue, 33 }  ,fill opacity=1 ][line width=1.5]  (323,250.25) .. controls (323,246.52) and (326.13,243.5) .. (330,243.5) .. controls (333.87,243.5) and (337,246.52) .. (337,250.25) .. controls (337,253.98) and (333.87,257) .. (330,257) .. controls (326.13,257) and (323,253.98) .. (323,250.25) -- cycle ; \draw  [line width=1.5]  (325.05,245.48) -- (334.95,255.02) ; \draw  [line width=1.5]  (334.95,245.48) -- (325.05,255.02) ;
\draw  [fill={rgb, 255:red, 126; green, 211; blue, 33 }  ,fill opacity=1 ][line width=1.5]  (434,250.25) .. controls (434,246.52) and (437.13,243.5) .. (441,243.5) .. controls (444.87,243.5) and (448,246.52) .. (448,250.25) .. controls (448,253.98) and (444.87,257) .. (441,257) .. controls (437.13,257) and (434,253.98) .. (434,250.25) -- cycle ; \draw  [line width=1.5]  (436.05,245.48) -- (445.95,255.02) ; \draw  [line width=1.5]  (445.95,245.48) -- (436.05,255.02) ;
\draw  [color={rgb, 255:red, 126; green, 211; blue, 33 }  ,draw opacity=1 ][line width=1.5]  (326.49,226.45) -- (334.51,215.55)(324.99,216.95) -- (336.01,225.05) ;
\draw  [color={rgb, 255:red, 126; green, 211; blue, 33 }  ,draw opacity=1 ][line width=1.5]  (436.29,225.63) -- (444.71,215.37)(435.79,216.63) -- (445.21,224.37) ;
\draw  [color={rgb, 255:red, 74; green, 144; blue, 226 }  ,draw opacity=1 ][line width=1.5]  (246.21,224.93) -- (254.56,215.55)(245.83,216.19) -- (254.94,224.3) ;
\draw  [color={rgb, 255:red, 74; green, 144; blue, 226 }  ,draw opacity=1 ][line width=1.5]  (526.45,225.48) -- (534.33,216.16)(526.06,217.16) -- (534.72,224.48) ;
\draw [color={rgb, 255:red, 189; green, 16; blue, 224 }  ,draw opacity=1 ][line width=1.5]    (126,204.5) -- (585,203.5) ;
\draw  [draw opacity=0][line width=1.5]  (126,204.5) .. controls (126.18,194.43) and (127.77,184.36) .. (130.87,174.46) .. controls (134.46,163.04) and (139.83,152.58) .. (146.68,143.22) -- (278.75,220.82) -- cycle ; \draw  [color={rgb, 255:red, 189; green, 16; blue, 224 }  ,draw opacity=1 ][line width=1.5]  (126,204.5) .. controls (126.18,194.43) and (127.77,184.36) .. (130.87,174.46) .. controls (134.46,163.04) and (139.83,152.58) .. (146.68,143.22) ;  
\draw  [draw opacity=0][line width=1.5]  (557.38,130.02) .. controls (577.18,149.12) and (586.56,176.02) .. (585,203.5) -- (486.02,206.2) -- cycle ; \draw  [color={rgb, 255:red, 189; green, 16; blue, 224 }  ,draw opacity=1 ][line width=1.5]  (557.38,130.02) .. controls (577.18,149.12) and (586.56,176.02) .. (585,203.5) ;  
\draw  [color={rgb, 255:red, 189; green, 16; blue, 224 }  ,draw opacity=1 ][line width=1.5]  (384,198.5) -- (394,204) -- (384,209.5) ;
\draw  [color={rgb, 255:red, 189; green, 16; blue, 224 }  ,draw opacity=1 ][line width=1.5]  (135.17,178.11) -- (127.46,183) -- (125.1,174.18) ;
\draw  [color={rgb, 255:red, 189; green, 16; blue, 224 }  ,draw opacity=1 ][line width=1.5]  (576.41,173.17) -- (579.94,164.83) -- (587,170.5) ;

\draw (266.5,274.25) node   [align=left] {\begin{minipage}[lt]{17pt}\setlength\topsep{0pt}
{\scalebox{1}{$\boldsymbol{i\delta}$}}
\end{minipage}};
\draw (320,270) node [anchor=north west][inner sep=0.75pt]   [align=left] {\scalebox{1}{$\boldsymbol{(a_{+}+i\delta)}$}};
\draw (429,269) node [anchor=north west][inner sep=0.75pt]   [align=left] {\scalebox{1}{$\boldsymbol{(a_{-}+i\delta)}$}};
\draw (518,268) node [anchor=north west][inner sep=0.75pt]   [align=left] {\scalebox{1}{$\boldsymbol{(1+i\delta)}$}};
\draw (99,120) node [anchor=north west][inner sep=0.75pt]   [align=left] { \scalebox{1}{$\boldsymbol{\lfloor \xi}$}};

\end{tikzpicture}
 \caption{Branch points of the $\xi$-integrand for $1<\zeta<\infty$ and our choice of the $\xi$-contour}
 \label{fifth_KLT}
\end{figure}

\end{document}